\documentclass[11pt]{article}
\usepackage[utf8]{inputenc}
\usepackage[left=2.54cm,right=2.54cm,top=2.54cm,bottom=2.54cm]{geometry}
\usepackage{multicol}
\usepackage{envmath}
\usepackage{graphicx}
\usepackage{multirow}
\usepackage{amsmath}
\usepackage{amsbsy}
\usepackage{amssymb}
\usepackage{subcaption}
\usepackage{graphicx}
\usepackage{siunitx}
\usepackage{float}
\usepackage{authblk}
\usepackage{orcidlink}
\providecommand{\keywords}[1]{\textbf{\textit{Keywords---}} #1}
\usepackage{url}
\usepackage{hyperref}
\usepackage{lipsum}
\usepackage{tikz,xcolor,hyperref}
\usepackage[
backend=biber,
style=numeric,
sorting=none
]{biblatex}

\definecolor{lime}{HTML}{A6CE39}
\DeclareRobustCommand{\orcidicon}{%
	\begin{tikzpicture}
	\draw[lime, fill=lime] (0,0) 
	circle [radius=0.16] 
	node[white] {{\fontfamily{qag}\selectfont \tiny ID}};
	\draw[white, fill=white] (-0.0625,0.095) 
	circle [radius=0.007];
	\end{tikzpicture}
	\hspace{-2mm}
}

\foreach \x in {A, ..., Z}{%
	\expandafter\xdef\csname orcid\x\endcsname{\noexpand\href{https://orcid.org/\csname orcidauthor\x\endcsname}{\noexpand\orcidicon}}
}

\title{\vspace{-1.5cm} \Huge Resonant behavior and stability of a linear three-mirror cavity}

\author[,1]{\large Paul Stevens \orcidA{} \thanks{Corresponding author: \texttt{paul.stevens@ijclab.in2p3.fr}}}
\author[,1]{Vincent Loriette \orcidF{} \thanks{Corresponding author: \texttt{vincent.loriette@ijclab.in2p3.fr}}}
\author[1]{Manuel Andia \orcidB{}}
\author[1]{François Glotin \orcidG{}}
\author[1]{Angélique Lartaux-Vollard \orcidC{}}
\author[1]{Nicolas Leroy \orcidD{}}
\author[1]{Aymeric van de Walle \orcidE{}}
    
\affil[1]{Université Paris-Saclay, CNRS/IN2P3, IJCLab, 91405 Orsay, France}

\date{\today}

\begin{document}

\maketitle

\title{\vspace{-0.5cm}}

\begin{abstract}
The implementation of Fabry-Perot cavities in gravitational-wave detectors has been pivotal to improving their sensitivity, allowing the observation of an increasing number of cosmological events with higher signal-to-noise ratio. Notably, Fabry-Perot cavities play a key role in the frequency-dependent squeezing technique, which provides a reduction of quantum noise over the whole observation frequency spectrum. In this context, linear three-mirror cavities could be of interest because of the additional control that they can provide.

In this paper, we develop a complete model to describe the stability behavior and the properties of transmitted and reflected fields of a linear three-mirror cavity aiming to be used for design purposes. In particular, simulations are carried out to show the evolution of the characteristic ``double-peak'' as a function of cavity parameters, which is one of the key features of this system. We show that the double-peak shape is almost freely adjustable, either in terms of spacing between maxima, their relative height and their intrinsic width. This is made possible by changing the mirrors' reflectivity coefficients and their spacing. However, the amount of achievable realistic configurations is limited by the stability conditions of the cavity. In particular, if the middle mirror is not close enough to the center of the cavity, it could be difficult to obtain a stable three-mirror cavity. Different geometries have been studied to obtain a stable cavity system.
\end{abstract}


\section{Introduction}

Fabry-Perot cavities are among the best-known optical systems. They are used in many fields, ranging from fundamental physics to engineering. They are formed by two parallel, partially-transmitting mirrors. The distance between them, as well as their transmission coefficients, determine the cavity's resonance properties. The resulting wavelength selectivity of this system is an essential feature of Fabry-Perot cavities which can be exploited in a wide variety of applications such as in lasers to generate a coherent beam or as a filter to select a specific frequency for telecom networks, or even in pressure or temperature sensors. They are also crucial in current and future generations of Gravitational-Wave Detectors (GWD) such as LIGO \cite{ligo}, Virgo \cite{virgo}, KAGRA \cite{kagra}, Einstein Telescope \cite{et_steering_committee_et_2020} or Cosmic Explorer \cite{cosmicExplorer}.

GWD are limited by several sources of noise, of which quantum noise is one of the most dominant. Despite the efficiency attained by frequency-dependent squeezing (FDS) \cite{ganapathy_broadband_2023}, the next generation of GWD such as Einstein Telescope -- Low Frequency will need different filtering systems to meet their own quantum noise targets \cite{et_steering_committee_et_2020}. For this purpose, different configurations of filter cavity are currently studied. Among them, linear three-mirror cavities could be particularly interesting because of the additional control they could offer for frequency-dependent squeezing. One of the key features of this system is the doubling of its resonance peak \cite{stadt, hogeveen, thuring, maggiore_tuning_2024}. However, the additional degrees of freedom of those cavities could make them more difficult to control.

In this paper, we develop a full model based on electromagnetic field propagation that we implement in a code to perform a non-exhaustive exploration of the optical properties and stability behavior of a general linear three-mirror cavity. Contrary to the study presented in \cite{thuring}, we are not elaborating here on the fundamental origin of the double peak, but on its shape variations as a function of the cavity configuration. We will not compare any further our solutions with the Fabry-Perot model but instead, we will focus on transmitted and reflected fields to understand how the design of the cavity impacts the resonance pattern. Some results obtained might not be relevant to our particular FDS problem but could motivate the use of this configuration for other goals in different optical systems.

\section{Three-mirror cavity optical behavior}

In the first part of this section, we propose an analytical model based on field propagation through the three-mirror cavity to explicitly express the global transmitted and reflected fields on cavity parameters. Then, in the second part, we will be interested in the evolution of the particular double peak resonance pattern as a function of those parameters by implementing the solutions of our model in a numerical simulation.

\subsection{Three-mirror cavity model}
\label{sect_cavityModel}

\begin{figure}[h!]
    \centering
    \includegraphics[width=0.7\textwidth]{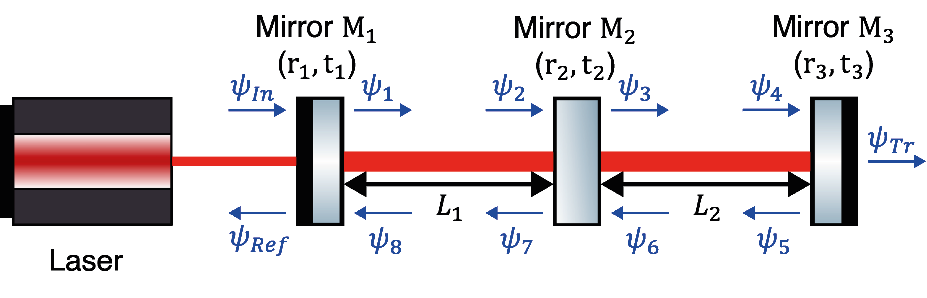}
    \caption{Three-mirror cavity scheme, the reflective surface of middle mirror can either be turn towards input or end mirror.}
    \label{fig_3MirCavScheme}
\end{figure}

To model the three-mirror cavity as a single optical system, we derive its overall transmission and reflection amplitude coefficients $r$ and $t$ by comparing the reflected and transmitted fields to a reference input field. On the three-mirror cavity scheme presented in Fig.~\ref{fig_3MirCavScheme}, we detail the position of each field that we use.

By following the path of a field entering the cavity and using the convention for which a reflected field is multiplied by the factor $r_\text{Surf}$ of a reflective surface while a transmitted field is multiplied by an imaginary transmission factor $it_\text{Surf}$ \cite{finesse}, we can construct the equation system (\ref{eq_systemFieldsEquation}). These equations depend on the following configuration parameters: $r_{1,2,3}$ and $t_{1,2,3}$ the amplitude reflection and transmission coefficients of input, middle and end mirror (respectively); $k=2\pi/\lambda$ the wave vector with $\lambda$ the wavelength of the input field; $L_{1,2}$ the distance between input and middle mirror or between middle and end mirror (i.e. first and second sub-cavities). Note that, in this model, we consider infinitely thin mirrors.

\begin{align}
    \begin{cases}
        \psi_1 = i t_1 \psi_\text{In} + r_1 \psi_{8} & \psi_2 = \psi_1 e^{-i k L_1} \\
        \psi_3 = i t_2 \psi_{2} + r_2 \psi_{6} & \psi_4 = \psi_3 e^{-i k L_2} \\
        \psi_5 = r_3 \psi_4 & \psi_6 = \psi_5 e^{-i k L_2} \\
        \psi_7 = i t_2 \psi_{6} + r_2 \psi_{2} & \psi_8 = \psi_7 e^{-i k L_1} \\
        \psi_\text{Ref} = i t_1 \psi_{8} + r_1 \psi_\text{In} & \psi_\text{Tr} = i t_3 \psi_{4}
    \end{cases}
    \label{eq_systemFieldsEquation}
\end{align}

From this cavity parameterization, we can derive the expression of reflection and transmission coefficients of the full three-mirror cavity $r = \frac{\psi_\text{Ref}}{\psi_\text{In}}$ and $t = \frac{\psi_\text{Tr}}{\psi_\text{In}}$ as a function of configuration parameters.

The reflection coefficient:
\begin{equation}
    r = \frac{\psi_\text{Ref}}{\psi_\text{In}} = \frac{i t_1 \psi_{8} + r_1 \psi_\text{In}}{\psi_\text{In}}
\end{equation}
can be rewritten in terms of $\psi_2$ and $\psi_6$:
\begin{equation}
    r = \frac{-t_1^2e^{-ikL_1}(it_2\psi_{6}+r_2\psi_{2})+r_1(\psi_{2}e^{ikL_1}-r_{1}e^{-ikL_1}(it_{2}\psi_{6}+r_{2}\psi_{2}))}{\psi_{2}e^{ikL_1}-r_1e^{-ikL_1}(it_2\psi_{6}+r_2\psi_{2}))}
    \label{eq_calculCoeffRef_step2}
\end{equation}

Then, expressing $\psi_{2}$ as a function of $\psi_{6}$:
\begin{equation}
    \psi_{2} = \frac{\psi_{6}}{it_{2}}\left(\frac{e^{2ikL_2}}{r_3}-r_2\right)
    \label{eq_calculCoeffRef_step3}
\end{equation}
and, finally, injecting Eq.~(\ref{eq_calculCoeffRef_step3}) in Eq.~(\ref{eq_calculCoeffRef_step2}), we obtain:
\begin{equation}
    r = \frac{r_{1}e^{2ik(L_1+L_2)}-r_1r_2r_3e^{2ikL_1}-r_2(r_1^2+t_1^2)e^{2ikL_2}+r_3(r_1^2+t_1^2)(r_2^2+t_2^2)}{e^{2ik(L_1+L_2)}-r_1r_2e^{2ikL_2}-r_2r_3e^{2ikL_1}+r_1r_3(r_2^2+t_2^2)}
    \label{eq_calculCoeffRef_result}
\end{equation}

In the same way, we can derive the expression of the transmission coefficient:
\begin{equation}
    t = \frac{-t_1t_2t_3e^{ik(L_1+L_2)}}{e^{2ik(L_1+L_2)}-r_1r_2e^{2ikL_2}-r_2r_3e^{2ikL_1}+r_1r_3(r_2^2+t_2^2)}
    \label{eq_calculCoeffTrans_result}
\end{equation}
These two equations are consistent with the results presented in \cite{stadt, thuring}.

From the above results, it is interesting to note that the global transmission and reflection coefficients of a three-mirror cavity are composed of both terms involving each two-mirror combination, comparable to those seen in Fabry-Perot reflection/transmission coefficients, and terms that involve all 3 mirrors. It follows that, in the general case, a three-mirror cavity cannot simply be understood as a series of two independent Fabry-Perot cavities nor a linear combination of intra-cavity fields of 3 cavities formed by each two-mirror combination, but a more complex system where the overall optical response depends on all the interdependencies and interactions between fields. Therefore, however simple the configuration with a third mirror may seem, we can understand from the above equations that a three-mirror cavity will have a fundamentally different behavior which cannot be understood as intuitively as for a Fabry-Perot configuration. However, this increase in complexity makes the three-mirror cavity system particularly interesting because of the specific resonance behavior it allows, as we will see in the following part. 

\subsection{Study of the splitting of the resonance}
\label{sect_doublePeakShape}

Based on coefficients solutions derived in section \ref{sect_cavityModel}, it is possible to analyze the optical response of a three-mirror cavity as a function of its configuration parameters.\\

\begin{figure}[h!]
    \centering
    \includegraphics[width=0.8\textwidth]{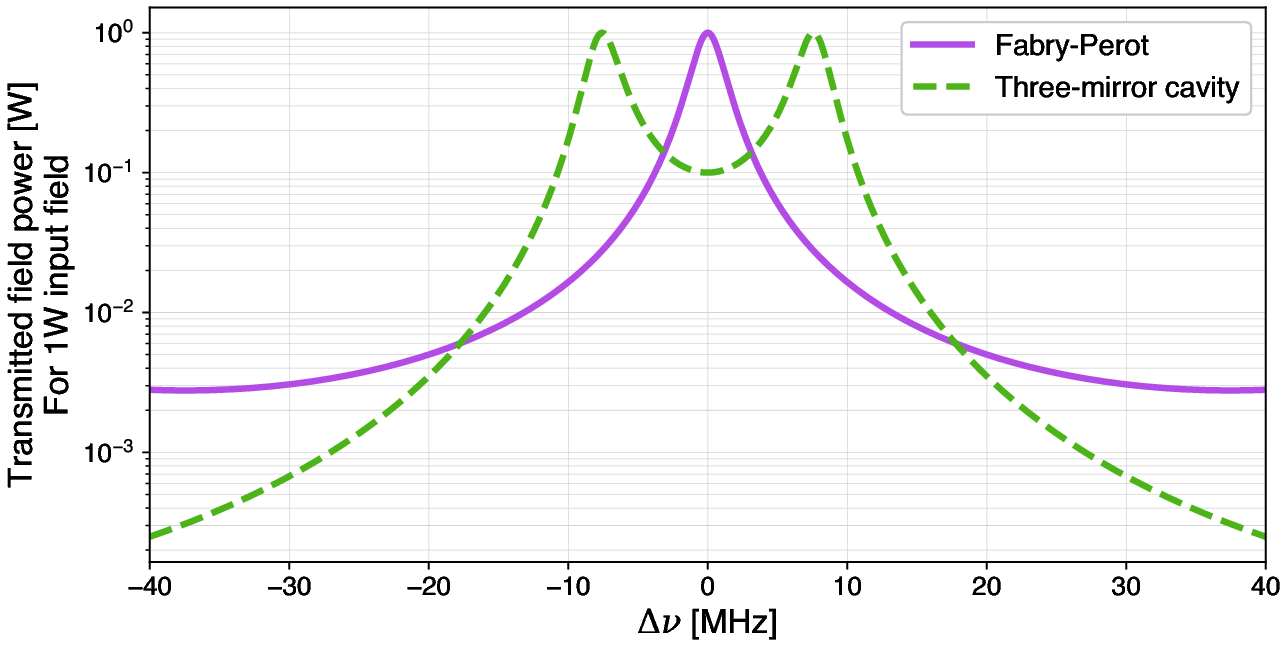}
    \caption{Transmitted field as a function of input-field frequency detuning, for an initial wavelength of $\lambda=\SI{1064}{nm}$. The Fabry-Perot cavity has the same overall length as the three-mirror configuration which is constituted by sub-cavities of length $L_1=L_2=\SI{1}{m}$. $\Delta \nu = 0$ corresponds to either the resonance of sub-cavities in the three-mirror configuration or the resonance of Fabry-Perot cavity.}
    \label{fig_FPvs3MC}
\end{figure}

For a Fabry-Perot configuration, an amplification of the field inside the cavity is expected for the resonance condition, i.e. cavity length being equal to an integer number of input-field half-wavelengths; by extension, scanning either the cavity length or the input field frequency would periodically lead to a maximization of transmitted power resulting in a peak distribution whose shape depends on cavity design and can be described by Airy functions \cite{ismail_fabry-perot_2016}. From a Fabry-Perot configuration, adding a third mirror between the two initial ones could result, as is represented on Fig.~\ref{fig_FPvs3MC}, in a resonant pattern that can no longer be described as a simple peak, as we will discuss in the last part of this section; there is a splitting of the initial single maxima \cite{stadt, thuring}. As we will see in this part, the shape of this ``double-peak'' depends on the cavity design and can be modulated in strongly different ways by changing the cavity configuration.

\subsubsection{Analytical calculation of the spacing between maxima}

From Eq.~(\ref{eq_calculCoeffTrans_result}), we can derive the analytical relation between three-mirror cavity parameters and the frequency spacing between each maximum of the double peak.
Squaring that equation and rewriting it in terms of $T_i=t_i^2$ leads to the expression of the global transmission factor:
\begin{equation}
    T=\frac{T_{1}T_{2}T_{3}}{UU^{*}}
\end{equation}
where:
\begin{equation}
    U=e^{2ik(L_1+L_2)}-r_1r_2e^{2ikL_2}-r_2r_3e^{2ikL_1}+r_1r_3(r_2^2+t_2^2)
\end{equation}
and $U^*$ its complex conjugate.

Then, using the definition of the wave vector $k$, the product $UU^*$ can be rewritten as a function of the detuning $\Delta\nu$ compared to the input field frequency:
\begin{equation}
    \begin{split}
        UU^* &= 1+R_{1}R_{2}+R_{2}R_{3}+R_{1}R_{3}(r_2^2+t_2^2)^2\\
        & -2(r_{1}r_{2}+r_{1}r_{2}R_{3}(r_2^2+t_2^2))\cos\left(\frac{4\pi}{c}\Delta\nu L_1\right)\\
        & -2(r_{2}r_{3}+R_{1}r_{2}r_{3}(r_2^2+t_2^2))\cos\left(\frac{4\pi}{c}\Delta\nu L_2\right)\\
        & +2r_{1}R_{2}r_{3}\cos\left(\frac{4\pi}{c}\Delta\nu(L_1-L_2)\right)\\
        & +2r_{1}r_{3}(r_2^2+t_2^2)\cos\left(\frac{4\pi}{c}\Delta\nu(L_1+L_2)\right)
    \end{split}
\end{equation}
where $R_i=r_i^2$.

To find at which frequency detuning corresponds the maxima as well as the minima between them, we should solve $dT/d(\Delta\nu)=0$. However, as the numerator in $T$ is constant, the problem boils down to the search of solutions of $d(UU^*)/d(\Delta\nu)=0$, where: 
\begin{equation}
    \begin{split}
        \frac{d(UU^*)}{d(\Delta\nu)} &= 2(r_{1}r_{2}+r_{1}r_{2}R_{3}(r_2^2+t_2^2))\frac{4\pi}{c}L_1\sin\left(\frac{4\pi}{c}\Delta\nu L_1\right)\\
        & +2(r_{2}r_{3}+R_{1}r_{2}r_{3}(r_2^2+t_2^2))\frac{4\pi}{c}L_2\sin\left(\frac{4\pi}{c}\Delta\nu L_2\right)\\
        & -2r_{1}R_{2}r_{3}\frac{4\pi}{c}(L_1-L_2)\sin\left(\frac{4\pi}{c}\Delta\nu(L_1-L_2)\right)\\
        & -2r_{1}r_{3}(r_2^2+t_2^2)\frac{4\pi}{c}(L_1+L_2)\sin\left(\frac{4\pi}{c}\Delta\nu(L_1+L_2)\right)
    \end{split}
    \label{eq_maxSpacing_result}
\end{equation}

This latter equation involves terms that depend on each part of the cavity. As a result, in this general case, the spacing of the maxima cannot be linked to any particular cavity parameter.\\ 

From Eq.~(\ref{eq_maxSpacing_result}), we could do substantial simplifications by choosing the example of a symmetrical configuration where each sub-cavity have the same length. It follows from the condition $L_1=L_2=L$, that Eq.~(\ref{eq_maxSpacing_result}) can be rewritten:
\begin{equation}
    \begin{split}
        \frac{d(UU^{*})}{d(\Delta\nu)} &= \frac{8\pi}{c} L\sin\left(\frac{4\pi}{c}\Delta\nu L\right) \left[r_1r_2(1+R_3(r_2^2+t_2^2)) \vphantom{\cos\left(\frac{4\pi}{c}\right)} \right.\\
        & \left. +r_2r_3(1+R_1(r_2^2+t_2^2)) -4r_1r_3(r_2^2+t_2^2)\cos\left(\frac{4\pi}{c}\Delta\nu L\right) \right] 
    \end{split}
\end{equation}

This new equation has 3 solutions. The more trivial one is $\Delta\nu=0$: this lets us understand that, for a symmetrical configuration, regardless of the other cavity parameters, the position of the minima between the two maxima always remains the same and corresponds to the anti-resonance frequency of a Fabry-Perot cavity of the same overall length. The second and third solutions are found for frequencies:
\begin{equation}
    \Delta\nu_{\pm}= \pm \frac{c}{4\pi L}\arccos \left[\frac{r_1r_2(1+R_3(r_2^2+t_2^2))+r_2r_3(1+R_1(r_2^2+t_2^2))}{4r_1r_3(r_2^2+t_2^2)}\right]
    \label{eq_maxSpacing_simplified_result}
\end{equation}
From the latter solutions, which have the same absolute value, we can see that each maximum will always be equally spaced from $\Delta\nu=0$. As a result, the resonance patterns achievable for symmetrical configurations are always symmetrical with maxima spaced by a total frequency range $\Delta_\text{Max} = 2 \lvert \Delta\nu_\pm \rvert$ and centered on zero detuning.\\

Finally, by adding the condition of highly reflective mirrors  $R_1\approx\ R_2\approx\ R_3\approx\ 1$, Eq.~(\ref{eq_maxSpacing_simplified_result}) can be simplified even further:
\begin{equation}
    \Delta\nu_{\pm} \approx \pm\frac{t_{2}c}{4\pi L}
\end{equation}

This expression, which corresponds to the results presented in \cite{maggiore_tuning_2024}, highlights the important role played by the central mirror in highly-reflective symmetrical cavities.\\

\subsubsection{Simulations of resonance behavior as a function of cavity parameters}

To understand in a more general way the variations of the double peak shape as a function of configuration changes, we realized simulations for which we computed the three-mirror transmitted field as a function of input field frequency detuning and cavity-parameter variations. For all of the following Figures in this section, we started from the same initial meter scale configuration $L_1=L_2=\SI{1}{m}$, with reflectivities of mirrors being $R_1=R_2=R_3=0.9$. Note that the frequency detuning value $\Delta\nu=0$ corresponds to a configuration where each sub-cavity is microscopically tuned to be resonant for an input field of wavelength $\lambda=\SI{1064}{nm}$.

\begin{figure}[h!]
    \centering
    \begin{subfigure}[h!]{.49\linewidth}
        \includegraphics[width=\textwidth]{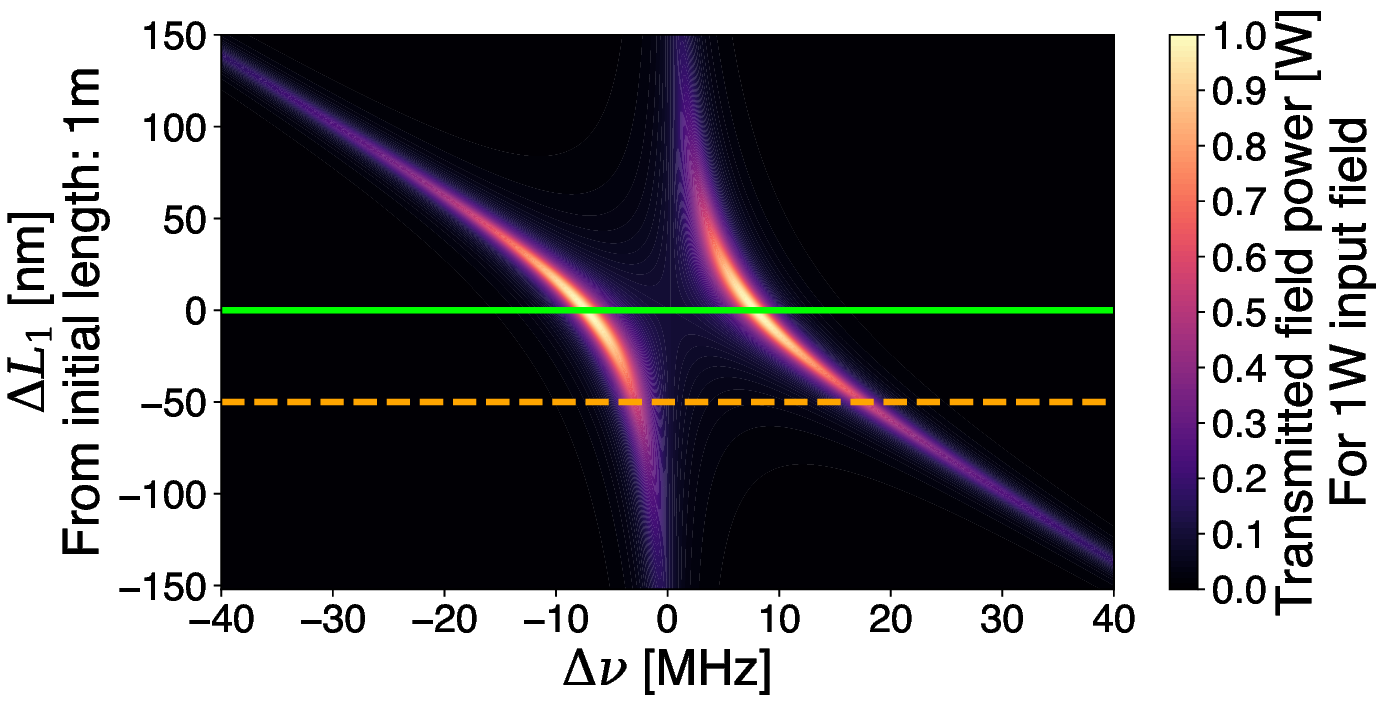}
        \caption{Transmitted field as a function of input-field detuning and detuning of the first sub-cavity's microscopic length. There is two asymptotes, which are vertical and diagonal, toward which each maxima tend as we increase the absolute value of $\Delta L_1$. The horizontal cuts at $\Delta L_1=\SI{0}{nm}$ and $\Delta L_1=\SI{-50}{nm}$ are represented on plot (\subref{fig_nuVSL_1Map_horizontalSlices}) opposite.}
        \label{fig_L1vsNu}
    \end{subfigure}
    \hfill
    \centering
    \begin{subfigure}[h!]{.49\linewidth}
        \includegraphics[width=\textwidth]{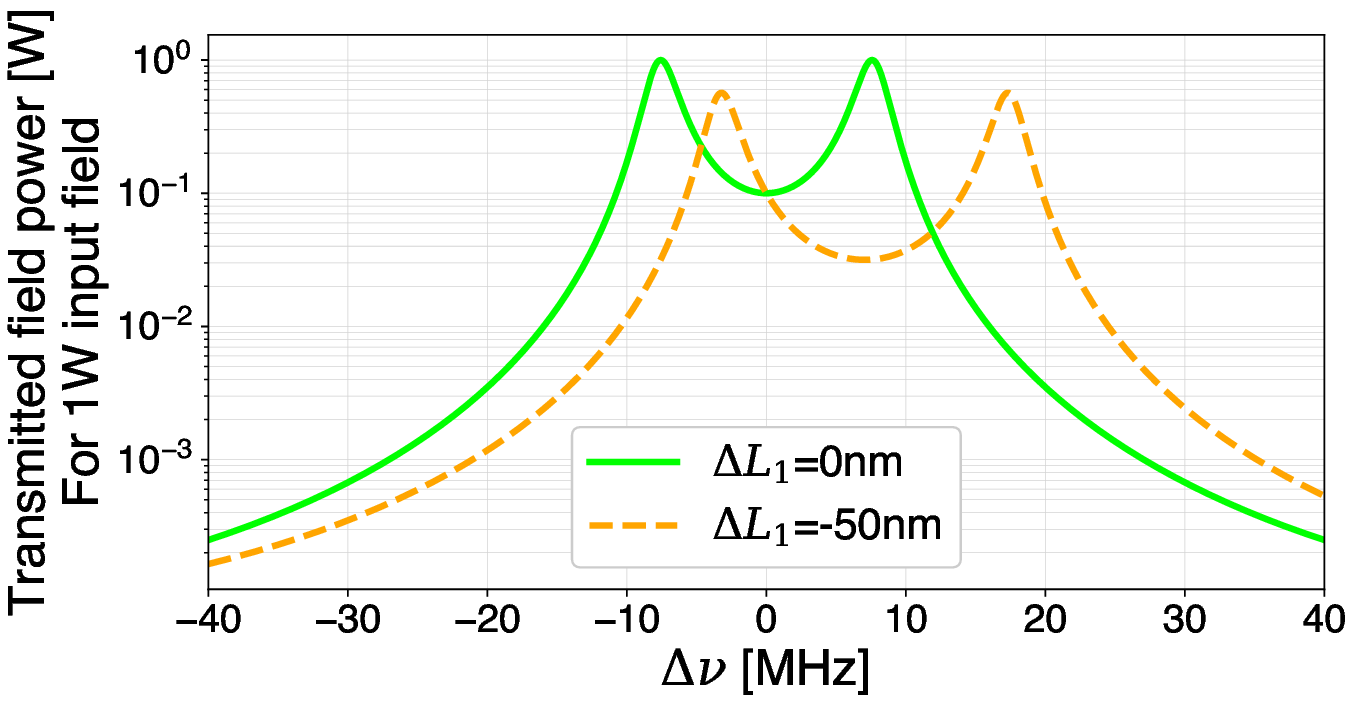}
        \caption{Transmitted field as a function of input-field detuning for two values of the microscopic length detuning of the first sub-cavity.}
        \label{fig_nuVSL_1Map_horizontalSlices}
    \end{subfigure}
    \caption{Position asymmetry of each double-peak maximum compared to $\Delta\nu=0$ for different values of the microscopic length detuning of the first sub-cavity.
    The reflectivities of mirrors are fixed to $R_1=R_2=R_3=0.9$, the first sub-cavity's length is modulated from its initial value $L_1=\SI{1}{m}$ while the second sub-cavity's length is kept fixed at $L_2=\SI{1}{m}$.}
    \label{fig_microscopicComparison}
\end{figure}

\begin{figure}[h!]
    \centering
    \includegraphics[width=0.8\textwidth]{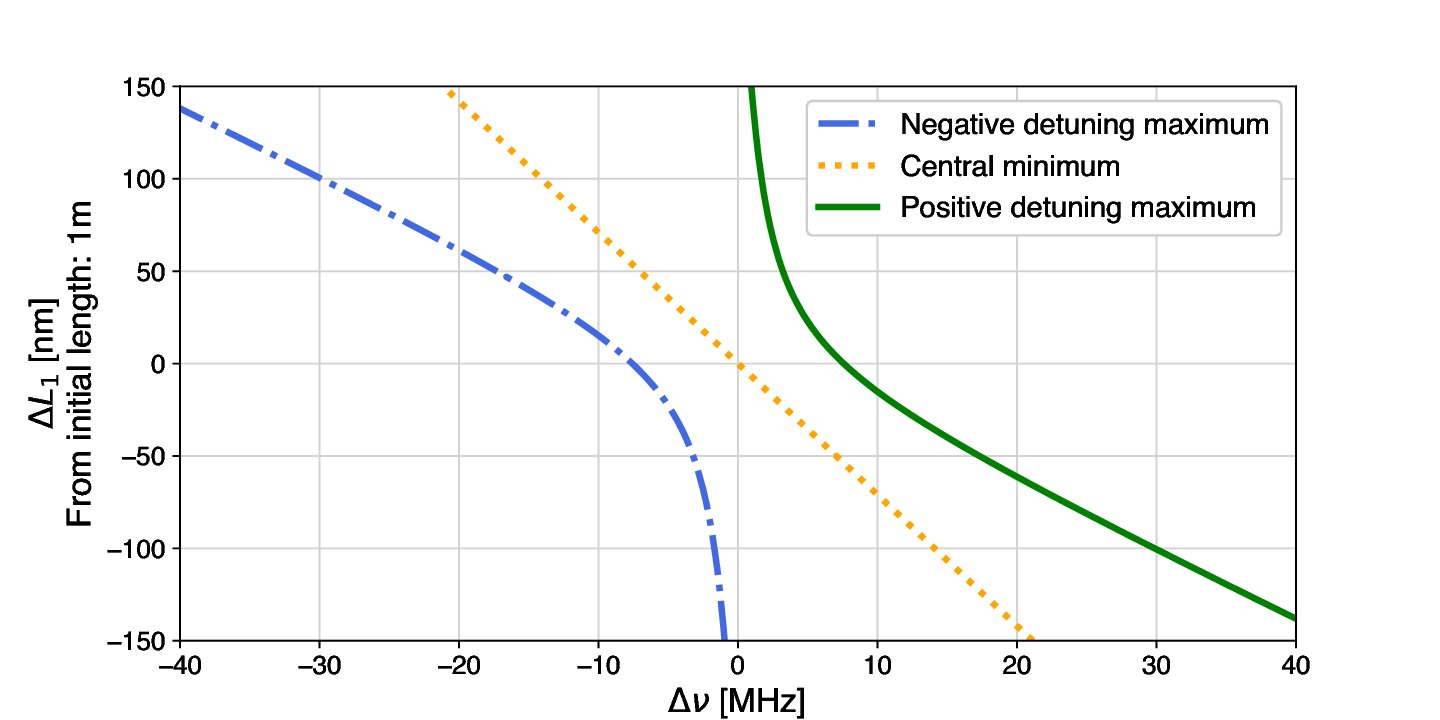}
    \caption{Position of the extrema of the reflection or transmission factors as a function of $L_1$ variation in nanometers. The first sub-cavity's length is modulated from its initial value $L_1=\SI{1}{m}$ while the second sub-cavity's length is kept fixed at $L_2=\SI{1}{m}$.}
    \label{fig_peaks_vs_dL1_v3}
\end{figure}

The parameters that can be set to design a three-mirror cavity are divided in two groups: the reflectivities of the mirrors and the lengths of the sub-cavities. Each category allows for changing in a different way the double-peak shape. In addition, macroscopic and microscopic variations of sub-cavities' lengths play different roles. As represented on Fig.~\ref{fig_microscopicComparison}, from a symmetrical 1m/1m configuration where both maxima are symmetrically distributed around zero detuning, which corresponds to a horizontal cut at $\Delta L_1 = 0$, introducing a microscopic asymmetry on the first sub-cavity would result in a different frequency shift for each maximum. As the asymmetry increases, one maximum shifts towards $\Delta\nu = 0$ while the other shifts away from it. The limit towards which each peak tends \cite{thuring} depends on the ratio between the lengths of the first and second sub-cavities, which is why we can see opposite behaviors on the upper and lower part of Fig.~\ref{fig_L1vsNu}. Note that all the positions of maxima and minima between them can be found by solving Eq.~(\ref{eq_maxSpacing_result}), as shown on Fig.~\ref{fig_peaks_vs_dL1_v3}.

\begin{figure}[h!]
    \centering
    \begin{subfigure}[t]{.49\linewidth}
        \includegraphics[width=\textwidth]{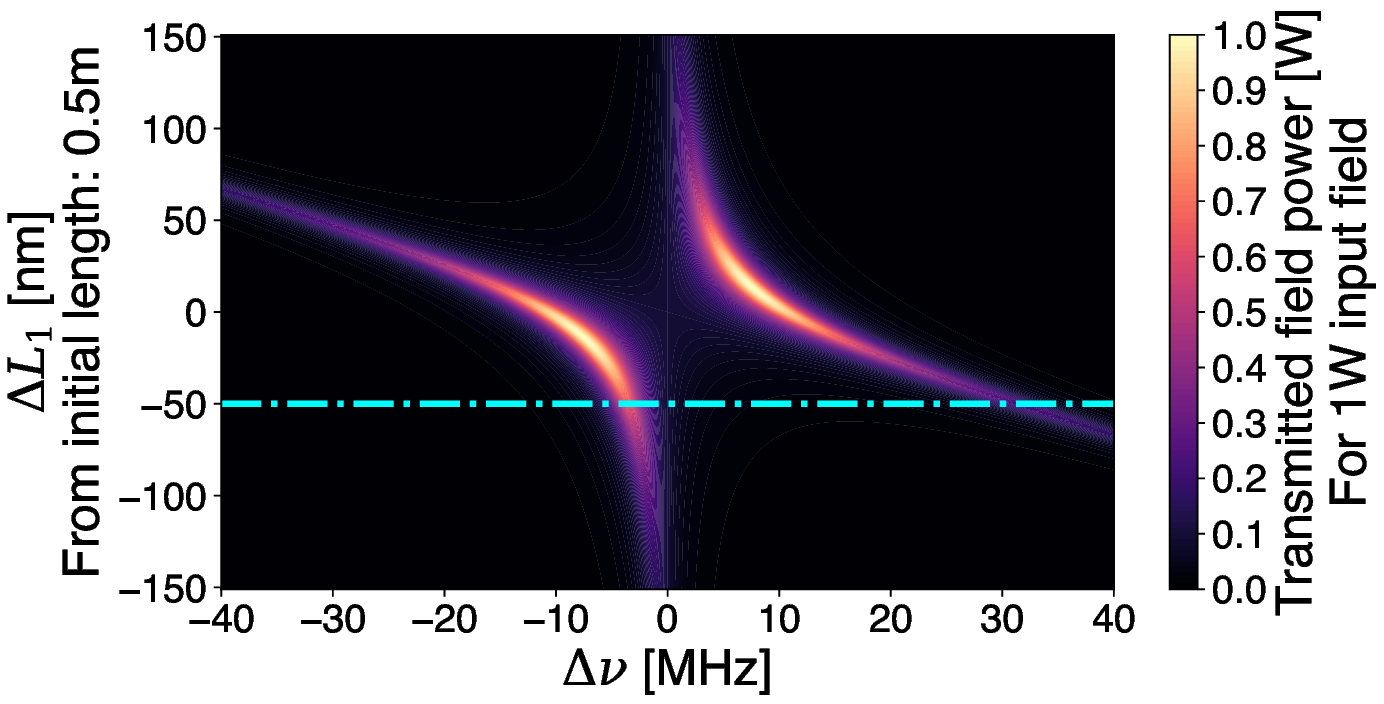}
        \caption{Initial lengths of the sub-cavities: $L_1=\SI{0.5}{m}$, $L_2=\SI{1}{m}$.}
        \label{fig_nuVSL_1Map_L=0.5}
    \end{subfigure}
    \hfill
    \begin{subfigure}[t]{.49\linewidth}
        \includegraphics[width=\textwidth]{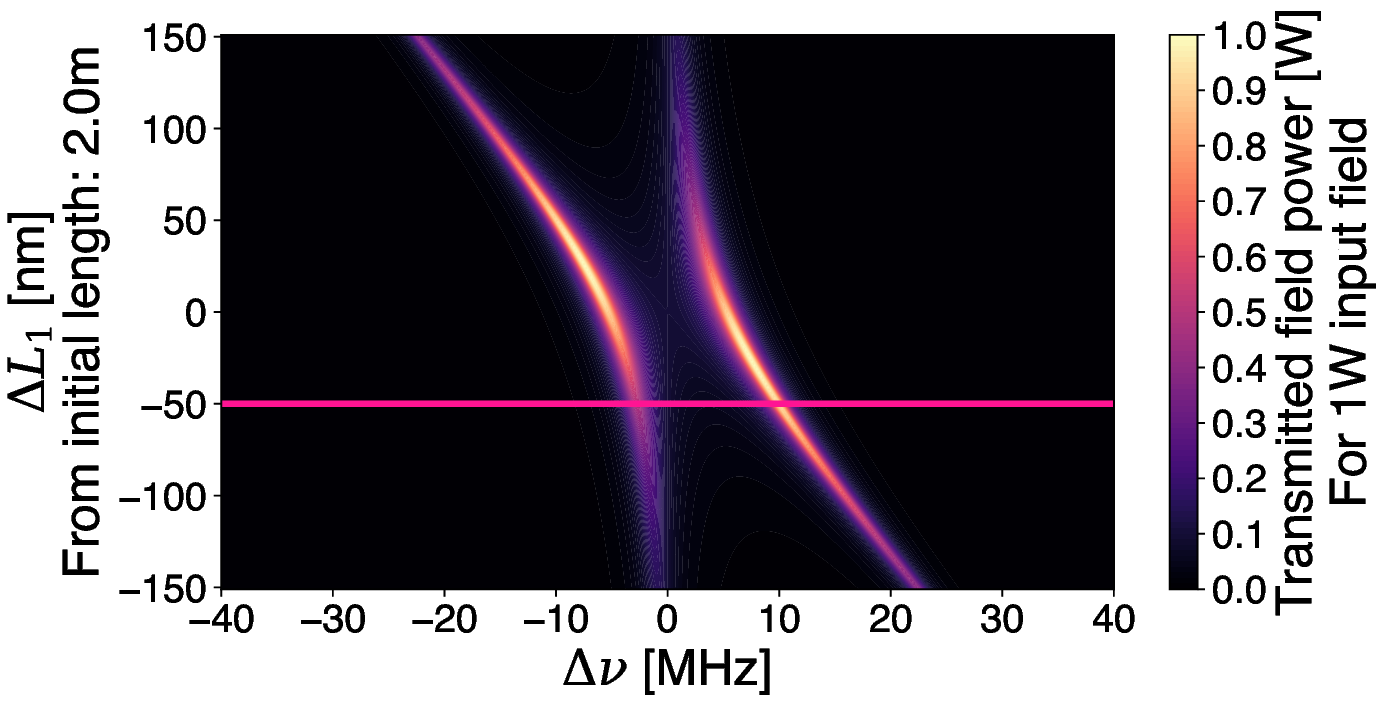}
        \caption{Initial lengths of the sub-cavities: $L_1=\SI{2}{m}$, $L_2=\SI{1}{m}$.}
        \label{fig_nuVSL_1Map_L=2}
    \end{subfigure}
    \hfill
    \begin{subfigure}[h!]{.49\linewidth}
        \includegraphics[width=\textwidth]{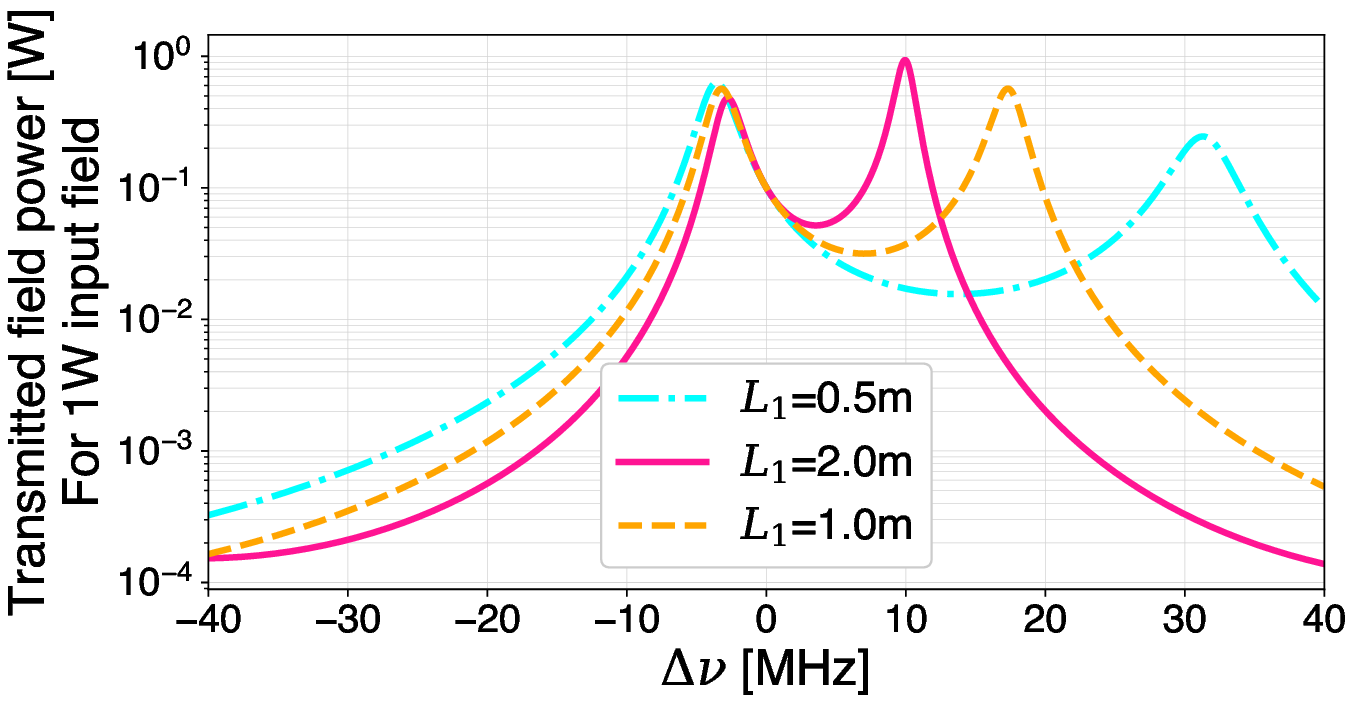}
        \caption{Transmitted field power as a function of detuning from (a) and (b) horizontal cuts. The symmetric configuration $L_1=L_2=\SI{1}{m}$ is represented in orange dashed line as a reference.}
        \label{fig_2_VS_0.5_horizCut}
    \end{subfigure}
    \caption{Illustration of peak drift rotation from initial sub-cavities lengths $L_1$ and $L_2$ of which the first is modulated by a microscopic detuning $\Delta L_1$. Mirrors reflectivity are set to $R_1=R_2=R_3=0.9$ and $L_2=\SI{1}{m}$ for each plot. On plots (a) and (b), the transmitted field is represented as a function of input field detuning and first sub-cavity microscopic length detuning. On plot (c), some specific $L_1/L_2$ ratio are represented for the same first sub-cavity detuning.}
    \label{fig_macroscopicComparison}
\end{figure}

In addition to the position asymmetry that can be induced by the microscopic tuning of one or the other sub-cavity's length, we can change the slope of the asymptote towards which the maxima moving further away from $\Delta\nu=0$ would tend. On Fig.~\ref{fig_macroscopicComparison}, we see that, for the same first sub-cavity microscopic length detuning, we obtain a double-peak whose right maximum drifts faster, and whose left maximum drifts slower, for a macroscopic length ratio $L_2=2L_1$ compared to the case where $L_1=2L_2$. In this way, by choosing a macroscopic asymmetry between sub-cavities, it is possible to control the frequency shift that a microscopic detuning would generate. Note that the microscopic detuning of a macroscopically asymmetric three-mirror cavity would also result in a change of relative height between maxima as we see on Fig.~\ref{fig_2_VS_0.5_horizCut}. This is a projection effect that originates from the drift of maximum transmitted power along the extremum position traces.

Although setting microscopic and macroscopic asymmetry between sub-cavities would result in an asymmetrical change of the positions of the maxima and a drift of the maximum transmitted power along extremum position traces, it is possible to act only on the latter by changing the ratio between the first and the third mirror reflectivity.

\begin{figure}[h!]
    \centering
    \begin{subfigure}[t]{.49\linewidth}
        \includegraphics[width=\textwidth]{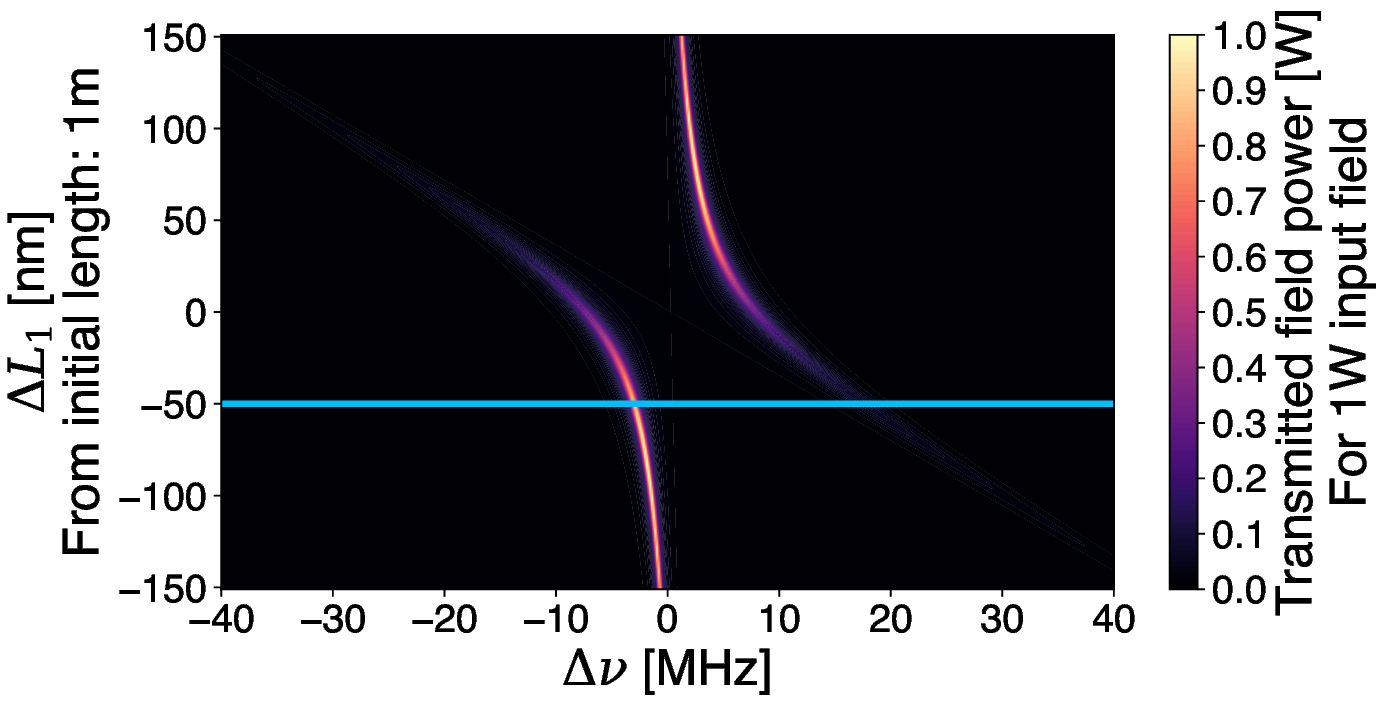}
        \caption{Mirror reflectivities are fixed to $R_1=R_2=0.9$ and $R_3=0.99$.}
        \label{fig_nuVSL_1Map_R1infR3}
    \end{subfigure}
    \hfill
    \begin{subfigure}[t]{.49\linewidth}
        \includegraphics[width=\textwidth]{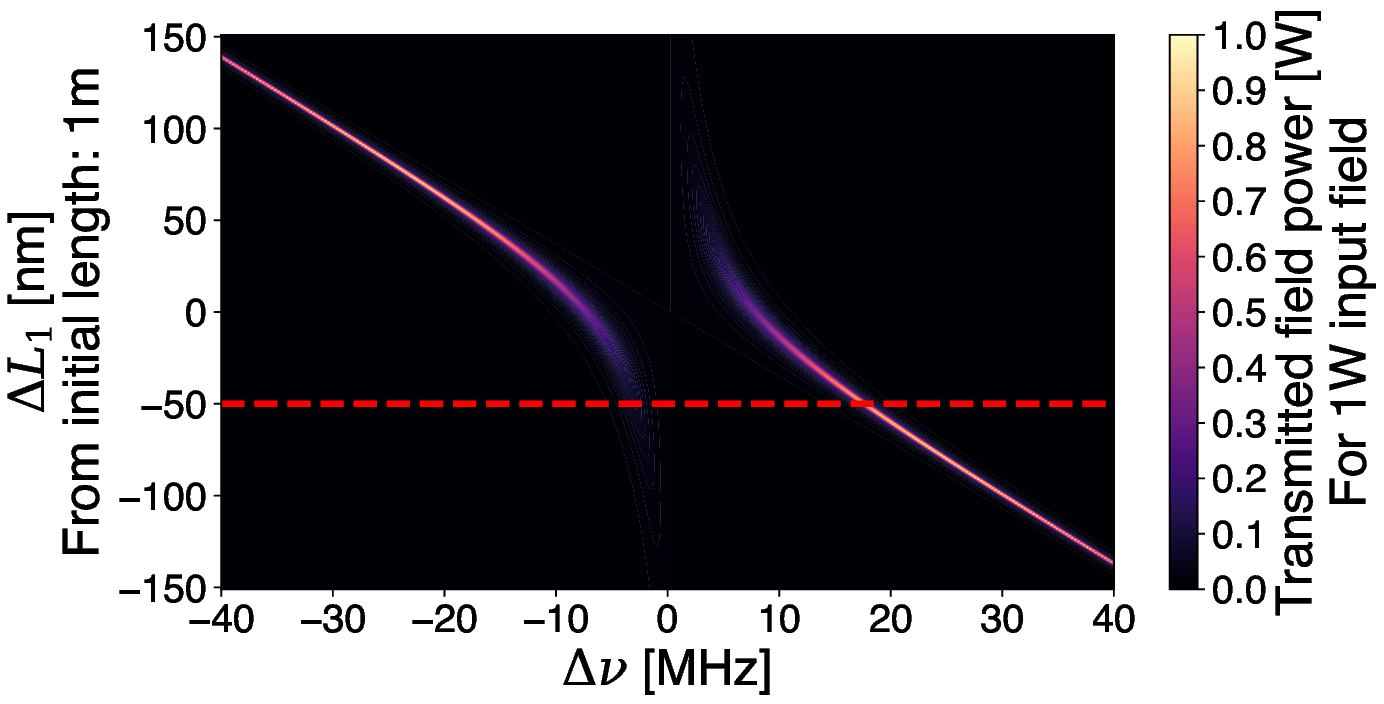}
        \caption{Mirror reflectivities are fixed to $R_1=0.99$ and $R_2=R_3=0.9$.}
        \label{fig_nuVSL_1Map_R1supR3}
    \end{subfigure}
    \hfill
    \begin{subfigure}[h!]{.49\linewidth}
        \includegraphics[width=\textwidth]{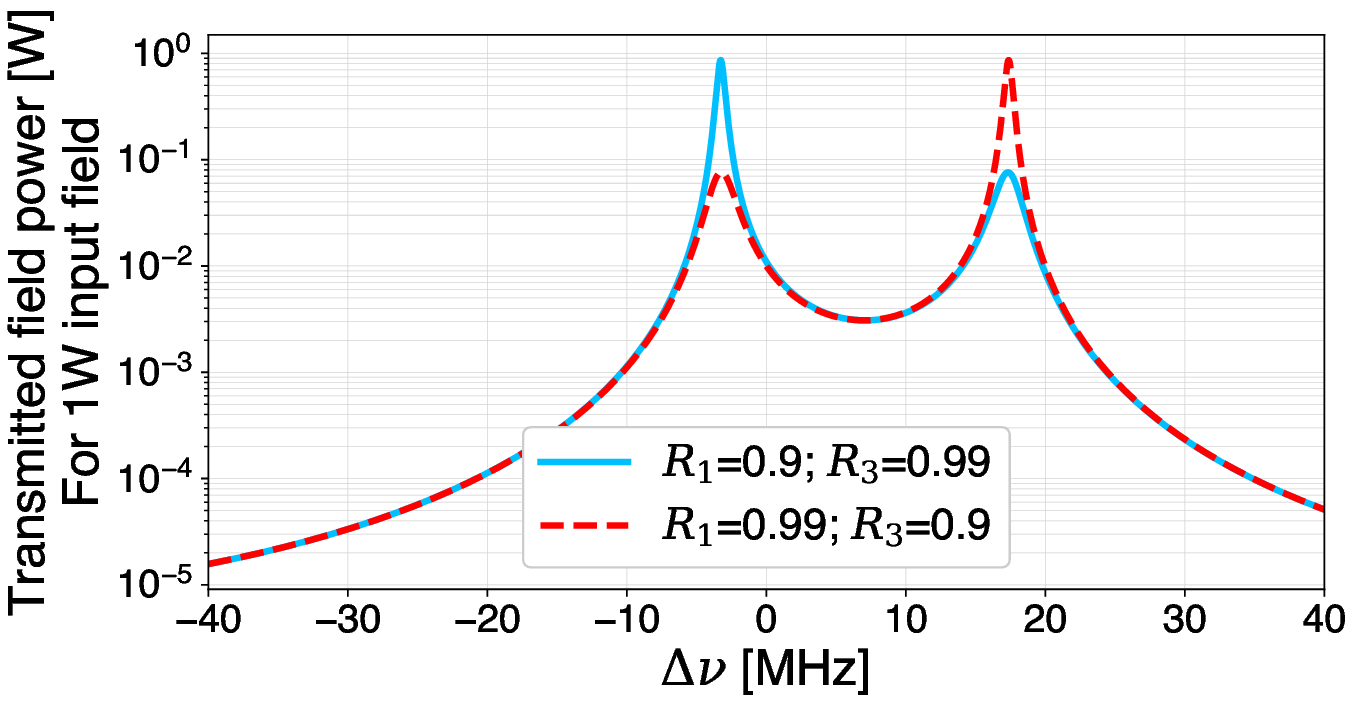}
        \caption{Transmitted field power as a function of detuning from (a) and (b) horizontal cuts.}
        \label{fig_nuVSL_1Map_R1vsR3}
    \end{subfigure}
    \caption{Power asymmetry between double-peak maxima as a function of the ratio $R_1/R_3$; (a) and (b) represent the transmitted field as a function of input field detuning and the first sub-cavity's microscopic length detuning. For each configuration, a horizontal cut at $\Delta L_1 = \SI{-50}{nm}$ is represented on plot (c).} 
\end{figure}

On Fig.~\ref{fig_nuVSL_1Map_R1infR3} and Fig.~\ref{fig_nuVSL_1Map_R1supR3}, we can see that setting the first and third mirror's reflectivity coefficients so that $R_1>R_3$ or $R_1<R_3$ lead to a drift of maximum power along resonance lines. On Fig.~\ref{fig_nuVSL_1Map_R1vsR3}, we can see that this result in variation of height ratio between maxima, which, for this particular case, strictly invert the resonance pattern, as we invert $R_1$ and $R_3$. 

\begin{figure}[h!]
    \centering
    \begin{subfigure}[t]{.49\linewidth}
        \includegraphics[width=\textwidth]{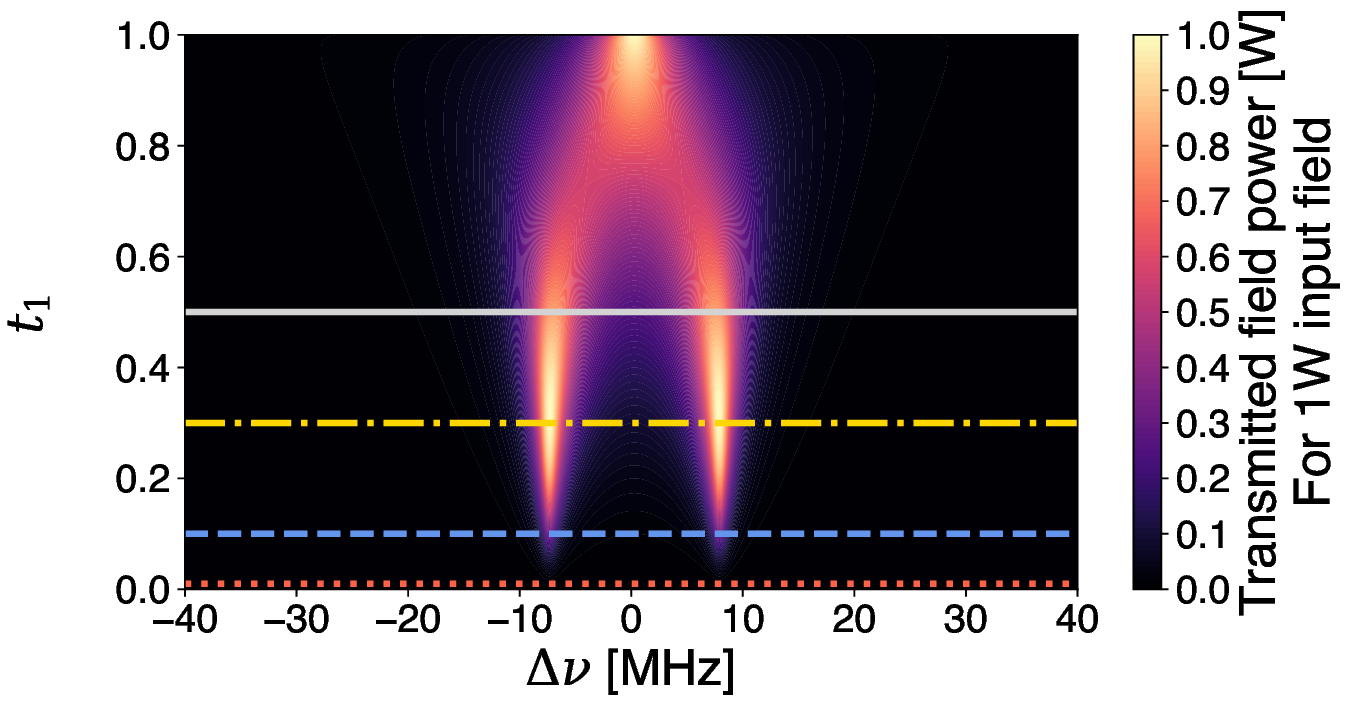}
        \caption{Transmitted field as a function of input field detuning and first mirror transmissivity. The horizontal cuts at $t_1=0.5$, $t_1=0.3$, $t_1=0.1$ and $t_1=0.01$ are represented on plot (\subref{fig_nuVScav_t_1Map_horizCuts}) opposite.}
        \label{fig_nuVScav_t_1Map_powerMap}
    \end{subfigure}
    \hfill
    \centering
    \begin{subfigure}[t]{.49\linewidth}
        \includegraphics[width=\textwidth]{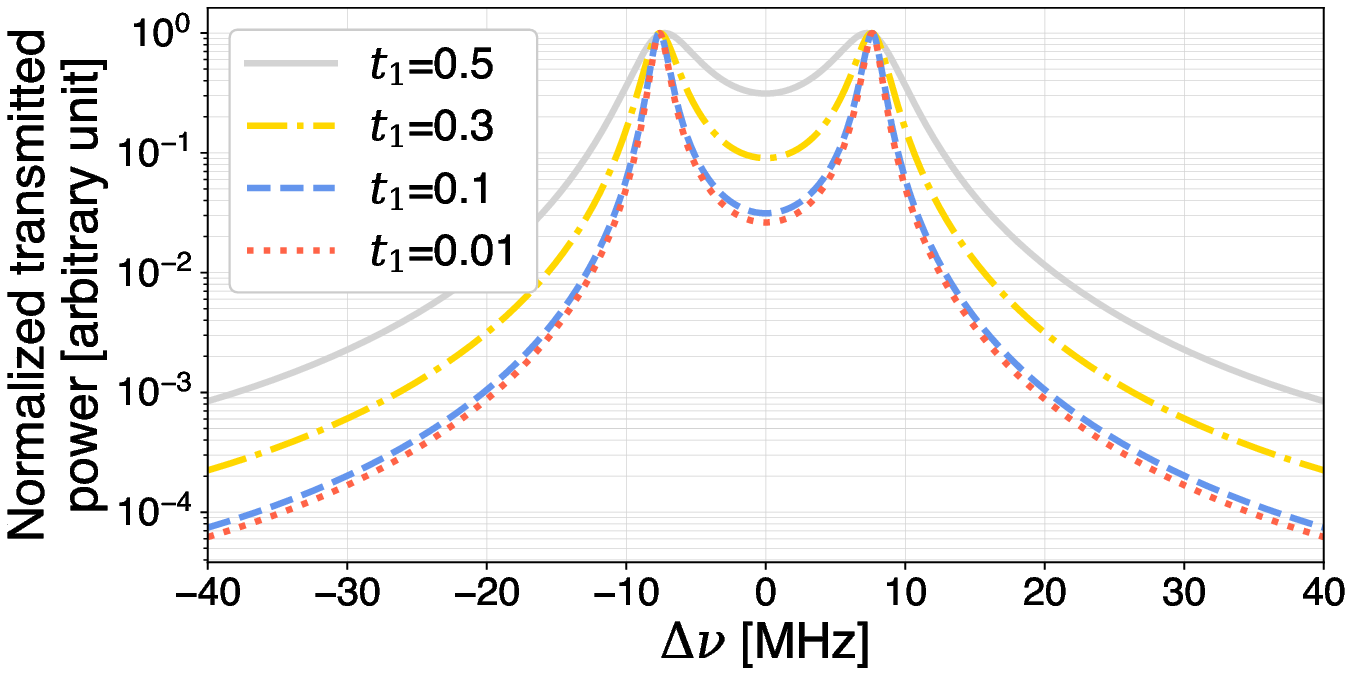}
        \caption{Normalized transmitted field power as a function of input field detuning for different values of the transmissivity coefficient of first mirror.}
        \label{fig_nuVScav_t_1Map_horizCuts}
    \end{subfigure}
    \caption{Transmitted field as a function of input mirror transmission $t_1$ and input field detuning. Other parameters are fixed to $R_2=R_3=0.9$ and $L_1=L_2=\SI{1}{m}$.}
\end{figure}

\begin{figure}[h!]
    \centering
    \includegraphics[width=0.8\textwidth]{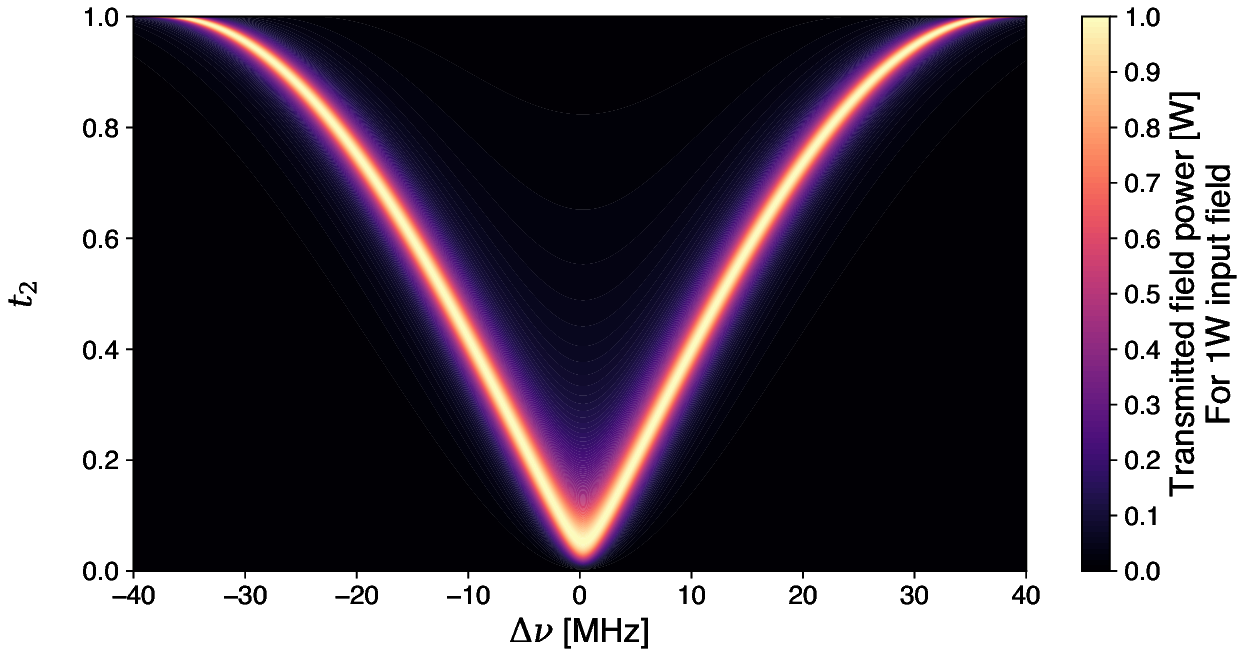}
    \caption{Transmitted field as a function of middle mirror transmissivity $t_2$ and input field detuning. Other parameters are fixed to $R_1=R_3=0.9$ and $L_1=L_2=\SI{1}{m}$. A horizontal cut at $t_2=1$ would correspond to a Fabry-Perot cavity of length \SI{2}{m} for which the free spectral range is about \SI{75}{MHz}.}
    \label{fig_nuVScav_t_2Map}
\end{figure}

On the other hand, the absolute reflectivity of the first and third mirrors are responsible for the intrinsic widths of the maxima. As can be seen on Fig.~\ref{fig_nuVScav_t_1Map_powerMap}, a fully transparent input mirror, which corresponds to a Fabry-Perot cavity constituted by the middle and third mirrors only, would lead, as expected, to a single peak. However, as the first mirror's transmissivity decreases, the initial peak splits in two maxima to form a double peak. By lowering even further the first mirror transmissivity, each peak becomes sharper but remains at the same frequency as represented on Fig.~\ref{fig_nuVScav_t_1Map_horizCuts}. On the contrary, the variation of middle mirror transmissivity leads to a different double-peak deformation. As shown on Fig.~\ref{fig_nuVScav_t_2Map}, starting from a configuration where $t_2=1$, which is also equivalent to a Fabry-Perot cavity, and decreasing its transmissivity would bring each maximum together until they completely merge for low $t_2$ values. Those results shows that the first/third mirrors and middle mirror play complementary roles on the relative position of the peaks and their intrinsic width. It is also interesting to note that, as we have set the microscopic length of each sub-cavity to be resonant, the equivalent Fabry-Perot cavity constituted by the first and third mirrors is anti-resonant. This can be seen on Fig.~\ref{fig_nuVScav_t_2Map} as the transmitted field power is minimized for input field detuning $\Delta \nu = 0$ at $t_2=1$.\\

\begin{figure}[h!]
    \centering
    \includegraphics[width=0.8\textwidth]{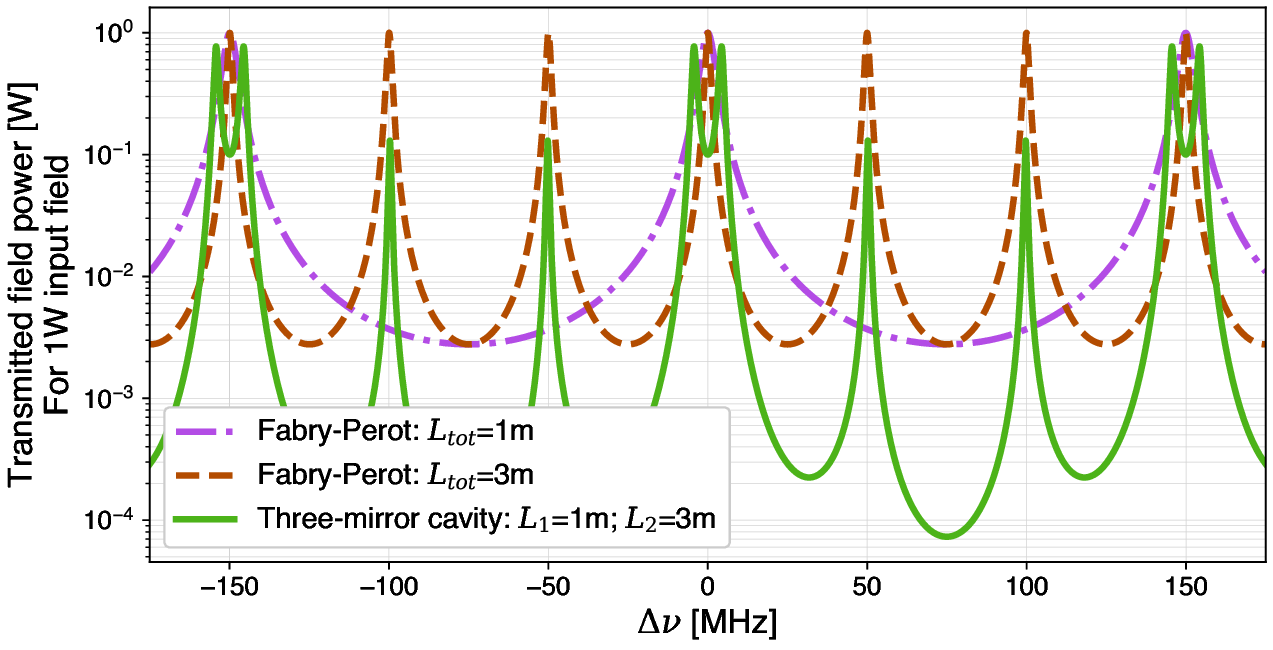}
    \caption{Transmitted field power as a function of input-field detuning, for a three-mirror configuration with sub-cavities of respective lengths $L_1=\SI{1}{m}$ and $L_2 = \SI{3}{m}$ (solid line), and two separate Fabry-Perot configurations with respective lengths $L_\text{tot} = \SI{1}{m} = L_1$ and $L_\text{tot} = \SI{3}{m} = L_2$ (dashed lines).}
    \label{fig_condApparitionDoublePic}
\end{figure}

As we have seen, the double peak shape can be modulated in strongly different ways thanks to the mirrors' reflectivities and sub-cavities' lengths, however it does not show off for all three-mirror cavity configuration. A necessary condition to obtain the double peak pattern is that each sub-cavity has to be close to its own resonance. In this way, the ratio between intrinsic sub-cavities' free spectral range governs the ratio between the number of double and single peaks \cite{thuring}. Moreover, as we can see in Fig.~\ref{fig_condApparitionDoublePic}, single peaks appearing between double peaks are highly distorted compared to those of a two-mirror Fabry-Perot cavity.

To conclude this part, from these simulations we show that the shape of the double peak is almost completely adaptable by designing a three-mirror cavity for which the double peak appearance conditions are met. However, the modulation of the double-peak shape goes beyond the scope of the design. Indeed, as the macroscopic distances separating the mirrors of an already implemented cavity could be microscopically detuned in real time, it could be possible to let the resonance behavior of the system be adaptable and not only fixed by design as is the case with Fabry-Perot cavities. Conversely, this important increase of versatility comes with the counterpart of more complex stability behavior, this is what we present in the next part.

\section{Stability}
\subsection{Preliminary considerations}
\label{sec:stabintro}
Up to this point, we have described the particular resonance behavior of a linear three-mirror cavity, but we simplified our approach by considering infinite flat mirrors. Still ignoring diffraction or other loss sources, we are now going to focus on stability behavior and how it dictates their geometries. The stability properties studied here are purely geometrical, so we will not consider the coupling by the intermediate mirror and restrict our study to the stability of two cavities sharing a common mirror and a common beam geometry.
In what follows we will thus describe a three-mirror cavity as made of
two half-cavities or equivalently two cavities. It is clear that those
two (half-)cavities are simple two-mirror Fabry-Perot cavities sharing
a common mirror.

\begin{figure}[h!]
    \centering
    \includegraphics[width=0.5\textwidth]{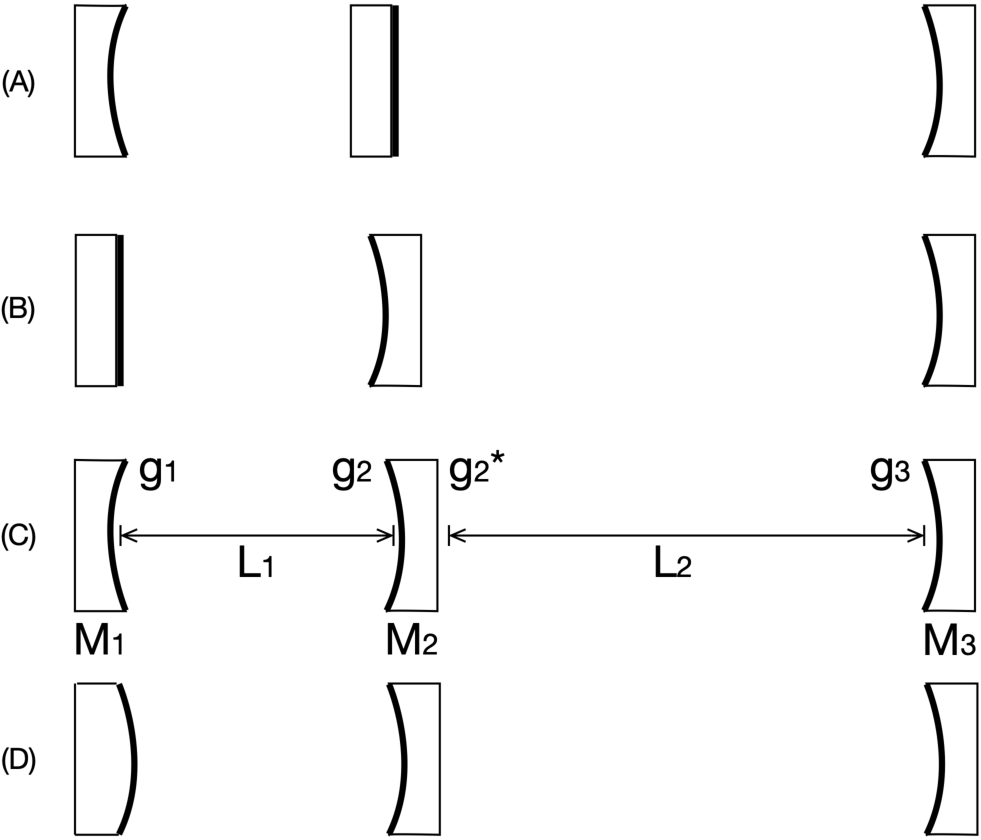}
    \caption{Three-mirror stable cavities. (A) Concave-plano-concave, (B) plano-concave
    half-cavity and convex-concave second half, (C) biconcave half-cavity
    and convex-concave second half-cavity, (D) two convex-concave half-cavities.
    \label{fig:Cavity_geometries}}
\end{figure}

A stable three-mirror cavity geometry can be of four
types, ignoring the fact that the substrate of the intermediate mirror can lie in either half-cavity. Those geometries are illustrated in Fig.~\ref{fig:Cavity_geometries}: (A) a flat central mirror and two concave end mirrors; (B) a flat end
mirror followed by a concave intermediate mirror and a concave end mirror; (C) a biconcave first cavity followed by a concave
end mirror, and (D) two convex-concave cavities. The second geometry (B) is in fact the same as the third one (C),
the flat-concave first cavity being one half of a biconcave cavity
where the flat mirror lies at the position of the beam waist. We will not study here the latter geometry (D) as it does not allow for designing cavities far from marginal stability. We are interested in three-mirror cavities where the two half-cavities have different lengths. One half-cavity of length $L_1$ is built with mirrors $M_1$ and $M_2$, and the second half-cavity of length $L_2$ is composed of mirrors $M_2$ and $M_3$. Each mirror $M_i$ has a radius of curvature $\rho_i$ and a flat surface anti-reflective backside.

The design of a stable three-mirror cavity is not as simple as it
may seem if one wants both halves of the cavity to be far from marginal
stability. The difficulty arises from the fact that a stable three-mirror
cavity cannot have a beam waist in both half-cavities, one of the
cavity imposes the beam geometry and the second half must then be
designed to be adapted to this imposed geometry. The stability criterion
of a two-mirror cavity can either be expressed (i) using the $g$ parameters as defined for example in \cite{ae_siegman_lasers_1986}: 
\begin{equation}
g_{1,2}=1-\frac{L}{\rho_{1,2}}
\end{equation}
where $L$ is the cavity length and $\rho_{1,2}$ are the mirrors' radii
of curvature, the relation:
\begin{equation}
0< g_{1}g_{2}<1
\end{equation}
ensuring the stability of the cavity, or (ii) using the Gouy phase accumulated in a cavity. If the Gouy phase is too small compared to the width
of the resonance peaks of the cavity, then high-order modes can easily
couple to the fundamental mode, making the cavity unstable. In fact,
both descriptions of stability contain the same information, as the
stability of a cavity essentially depends on the Rayleigh range of
the beam compared to the lengths of the cavities.

In all the following calculations we will use the $g$ parameters of the three mirrors. The stability of the first cavity will be described using $g_1$ and $g_2$ parameters, while the second cavity will be described using $g_2^*$ and $g_3$ parameters. Usually $g_2^* \neq g_2$ because of the effect of transmission by the anti-reflective back side of the intermediate mirror. Indeed, the second cavity ``sees'' the intermediate mirror $M_2$ as a mirror of radius of curvature $\rho_{2}^{*}$ that  is equal to:
\begin{equation}
\rho_{2}^{*}\approx\frac{-\rho_{2}}{n}
\end{equation}
using a first order approximation of Snell's law where $n$ is the index of refraction of mirror $M_2$
substrate.

In the process of designing a three-mirror non-symmetrical cavity, one half-cavity necessarily dictates the design of the other half.  We will show in Section \ref{sec:CPC} that in configuration (A) on Fig. \ref{fig:Cavity_geometries} of concave plano-concave cavity, whatever the cavity which is designed first, it is not possible to obtain a very stable second cavity if the length ratio $L_1/L_2$ is far from unity. However configuration (C) on Fig. \ref{fig:Cavity_geometries}, with a biconcave cavity and a convex-concave cavity, offers stable solutions for ratio $L_1/L_2$ farther from unity for some configurations as shown in Section \ref{sec:CCC}. Finally we will show in Section \ref{sec:lens-mirror} the impact on the overall cavity stability of designing the intermediate mirror as a lens-mirror component.

\subsection{Concave-plano-concave cavity}
\label{sec:CPC}

The flat central mirror has a flat anti-reflective side, that side lays inside effect of refraction at this anti-reflective side by assuming that the mirror thickness is much smaller than the beam's Rayleigh range so the beam radius of curvature at the anti-reflective side location is very large and its change has a small influence on the stability of the system.
Let us start by fixing the geometry of the first half-cavity by imposing its stability
factor. We choose to express it as:
\begin{equation}
g_{1}g_{2}=\frac{1}{2}+\varepsilon
\end{equation}
where $\varepsilon\in\left]-1/2,1/2\right[$.
The intermediate mirror
being flat, $g_{2}=1$ and we can express $\rho_{1}$ the radius of curvature
of the concave mirror forming the first cavity as:
\begin{equation}
\rho_{1}=\frac{2L_{1}}{1-2\varepsilon}
\end{equation}

The Rayleigh range $z_R$ depends on the beam waist size $w_0$ as $z_R=\pi w_0^2/\lambda$, so it depends on the cavity geometry. The relationship between the Rayleigh range, the input mirror radius of curvature $\rho_{1}$ and the first half-cavity length $L_{1}$ is \cite{ae_siegman_lasers_1986}:
\begin{equation}\label{eq:Rayleigh}
z_{R}^{2}=L_{1}\left(\rho_{1}-L_{1}\right)
\end{equation}
so:
\begin{equation}
z_{R}^{2}=L_{1}^{2}\frac{1+2\varepsilon}{1-2\varepsilon}
\end{equation}

This Rayleigh range value is the same in the second half-cavity because
both cavities share the same beam waist size at the same location.
The radius of curvature $\rho_{3}$ of the concave mirror forming the
second cavity of length $L_{2}$ is thus:
\begin{equation}
\rho_{3}=L_{2}+\frac{z_{R}^{2}}{L_{2}}=L_{2}+\frac{L_{1}^{2}}{L_{2}}\left(\frac{1+2\varepsilon}{1-2\varepsilon}\right)
\end{equation}

The two half-cavities have different lengths, $L_{2}=mL_{1}$ with
$m \in \mathbb{R^{+*}}$ $\rho_{3}$ can be expressed as:
\begin{equation}
\rho_{3}=L_{2}\left(1+\frac{1}{m^{2}}\left(\frac{1+2\varepsilon}{1-2\varepsilon}\right)\right)
\end{equation}
and finally the stability factor of the second half-cavity is:
\begin{equation}
g_{2}g_{3}=g_{3}=1-\frac{1}{1+\frac{1}{m^{2}}\left(\frac{1+2\varepsilon}{1-2\varepsilon}\right)}
\end{equation}

\begin{figure}[h!]
    \centering
    \includegraphics[width=0.8\textwidth]{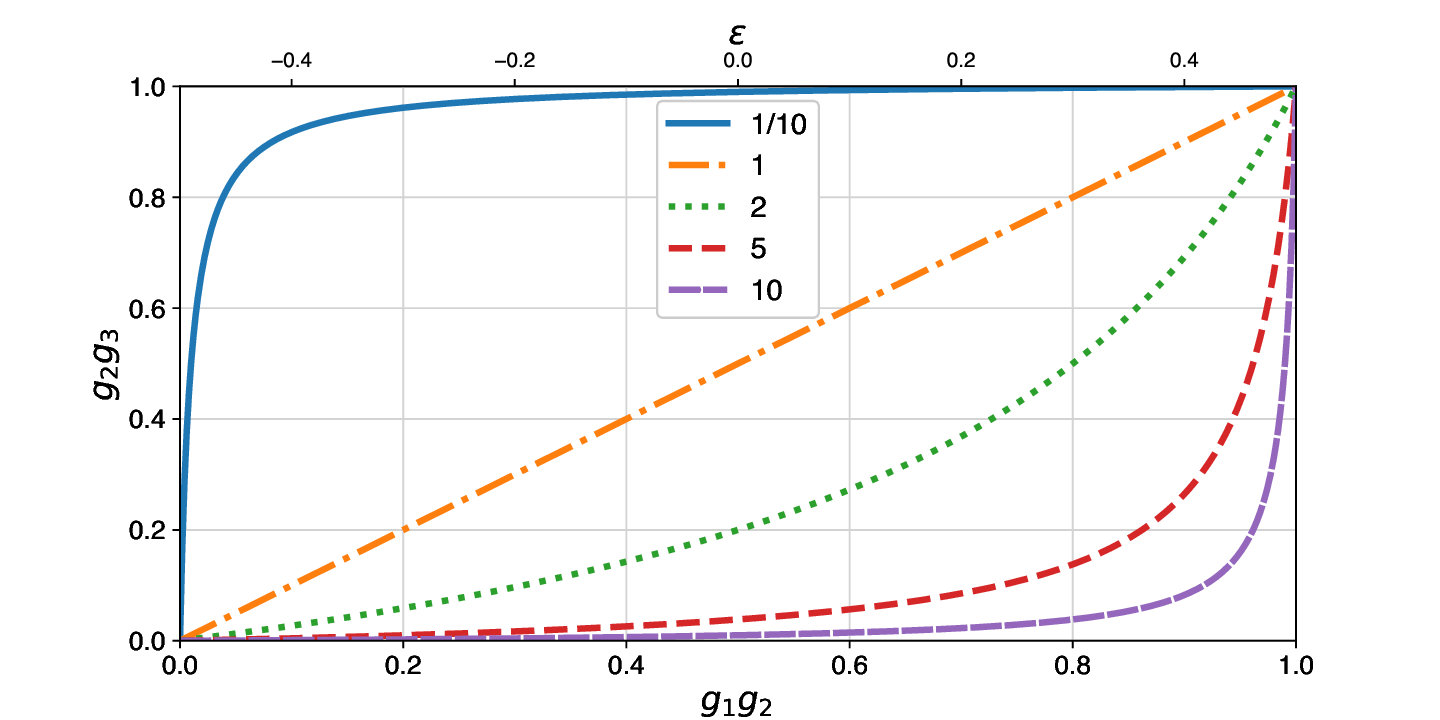}
    \caption{$g_{2}g_{3}$ vs. $g_{1}g_{2}$ for different values of $m$ in a
    concave-plano-concave cavity.
    \label{fig:g2g3vsg1g2}}
\end{figure}

Values of this stability factor for different values of $m=L_2/L_1$ are shown
on Fig.~\ref{fig:g2g3vsg1g2}.

We see that as soon as the value of $m$ becomes either large or small compared to one,
the stability factor $g_2 g_3$ rapidly either increases
towards 1 or decreases to 0, unless the value of $\varepsilon$ is
adapted to compensate this behavior. We see that as soon as $m$ exceeds a few units it becomes impossible to strongly stabilize both cavities simultaneously, so the concave-plano-concave geometry is not adapted to large length differences.

A simple explanation is this: the intermediate flat mirror imposes
its $g=1$. If the short cavity is designed first, it imposes a short
Rayleigh range. Because the long cavity length is large compared to
this Rayleigh range, its end mirror radius of curvature is close to
the cavity length, so the $g$ parameter of this mirror is always
close to zero and $g_{2}g_{3}$ of the long cavity is close to zero.
Conversely, if the long cavity is designed first then the $g_{1}$
parameter of the short cavity is close to one, and so is the $g_{1}g_{2}$
of the short cavity. It will appear later that building a three-mirror
cavity made of two stable cavities requires that the magnitude of
the intermediate mirror $g_{2}$ parameter seen from the long cavity
be large enough, either positive or negative, to compensate the necessarily
low value of the end mirror $g_{3}$ parameter.

\subsection{Biconcave cavity and convex-concave cavity}
\label{sec:CCC}

It is equivalent to build first the biconcave cavity and then the convex-concave one or the other way around. To our knowledge there is no well-known ``standard'' geometry
for a concave-convex cavity, thus we chose to design first the biconcave cavity of length $L_1$ and stability criterion $g_{1}g_{2}$ and then the convex-concave one of length $L_2$ and stability criterion $g_{2}^*g_{3}$ as depicted on configuration (C) of Fig. \ref{fig:Cavity_geometries} in Section \ref{sec:stabintro}.

In the simple case of a symmetric biconcave cavity of fixed length $L_1$ and given stability $g_{1}g_{2}$, there are 2 possible radii of curvature for its mirrors
\begin{equation}\label{eq:rho2}
\rho_1=\rho_2=\frac{L_1}{1\pm\sqrt{g_{1}g_{2}}}
\end{equation}
and conversely for a fixed radius of curvature $\rho_1=\rho_2$ and given stability $g_{1}g_{2}$, there are two possible cavity lengths.

Then for the convex-concave cavity of length $L_2$, a given radius of curvature $\rho_2$ of the intermediate mirror, with refractive index $n$ substrate inside the cavity, and a given stability $g_{2}^*g_{3}$ there is only one possible radius of curvature $\rho_3$
\begin{equation}\label{eq:rho3}
\rho_3=\frac{L_2}{1-\frac{\rho_2g_{2}^*g_{3}}{\rho_2+nL_2}}
\end{equation}
and conversely for a fixed radius of curvature $\rho_3$, a given radius of curvature $\rho_2$ of the intermediate mirror, with refractive index $n$ substrate inside the cavity, and a given stability $g_{2}^*g_{3}$ there is one possible positive cavity length $L_2$.

Now we geometrically have two choices: either build a long biconcave cavity or a short biconcave cavity. 

\subsubsection{Long biconcave cavity}

If we choose to build a biconcave long cavity, it happens to make it nearly impossible to obtain a stable short convex-concave cavity. Let us illustrate this point by studying a symmetric biconcave cavity of length $L_\text{long}$, with $\rho_\text{long}$ mirror radii of curvature and calculating its Rayleigh range as a function of the stability criterion $g_{1}g_{2}$. Equation (\ref{eq:Rayleigh}) becomes
\begin{equation}
z_{R}^{2}=\frac{L_\text{long}}{2}\left(\rho_\text{long}-\frac{L_\text{long}}{2}\right)
\end{equation}
and consequently the Rayleigh range is
\begin{equation}
z_{R}=\frac{L_\text{long}\sqrt{1-g_{1}g_{2}}}{2\left(1-\sqrt{g_{1}g_{2}}\right)}
\end{equation}

\begin{figure}[h!]
    \centering
    \includegraphics[width=0.8\textwidth]{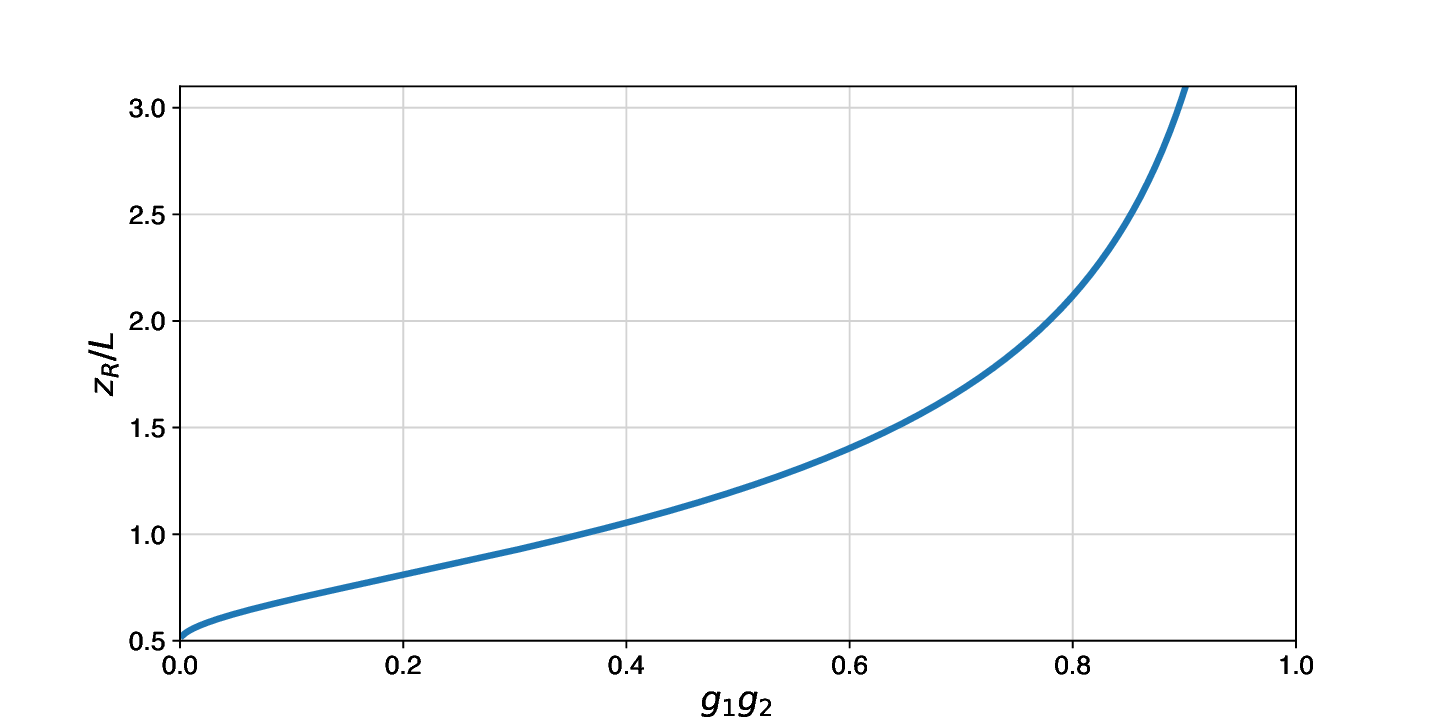}
    \caption{Rayleigh range in cavity length units of a symmetric biconcave cavity
    as a function of the stability criterion.
    \label{fig:zr_vs_g1g2}}
\end{figure}

The result is shown in Fig.~\ref{fig:zr_vs_g1g2}. As expected, for Rayleigh range below half of the cavity length, the cavity becomes unstable. Thus a long biconcave stable cavity imposes that the Rayleigh range be of the order of magnitude of its length, $z_R\approx L_\text{long}$. The beam waist being at the center of the long cavity, the Gouy phase accumulated in the short one,  $\phi_{Gs}$ is of the order of:
\begin{equation}
\phi_{Gs}=\arctan\left(\frac{L_\text{long}/2+L_\text{short}}{z_{R}}\right)-\arctan\left(\frac{L_\text{long}/2}{z_{R}}\right)
\end{equation}
\begin{equation}
\phi_{Gs}\approx\arctan\left(\frac{1}{2}+\frac{L_\text{short}}{L_\text{long}}\right)-\arctan\left(\frac{1}{2}\right)
\end{equation}
If one requires a Gouy phase larger than \ang{15} then the long cavity must not be longer than $2.7$ times the short cavity, which strongly limits the leeway to choose a length ratio. 

The same conclusion can be obtained by inspecting the $g_{2}^*g_{3}$ value of the short cavity. As the long cavity is designed to be stable, then
\begin{equation}
g_{2}^*g_{3}\approx\left(1+\frac{L_\text{short}}{L_\text{long}}\right)\times\left(1-\frac{L_\text{short}}{L_\text{long}}\right)
\end{equation}
which brings us to the same conclusion we drew considering the stability from a Gouy phase point of view.

\begin{figure}[h!]
    \centering
    \begin{subfigure}[h!]{.49\linewidth}
        \includegraphics[width=\textwidth]{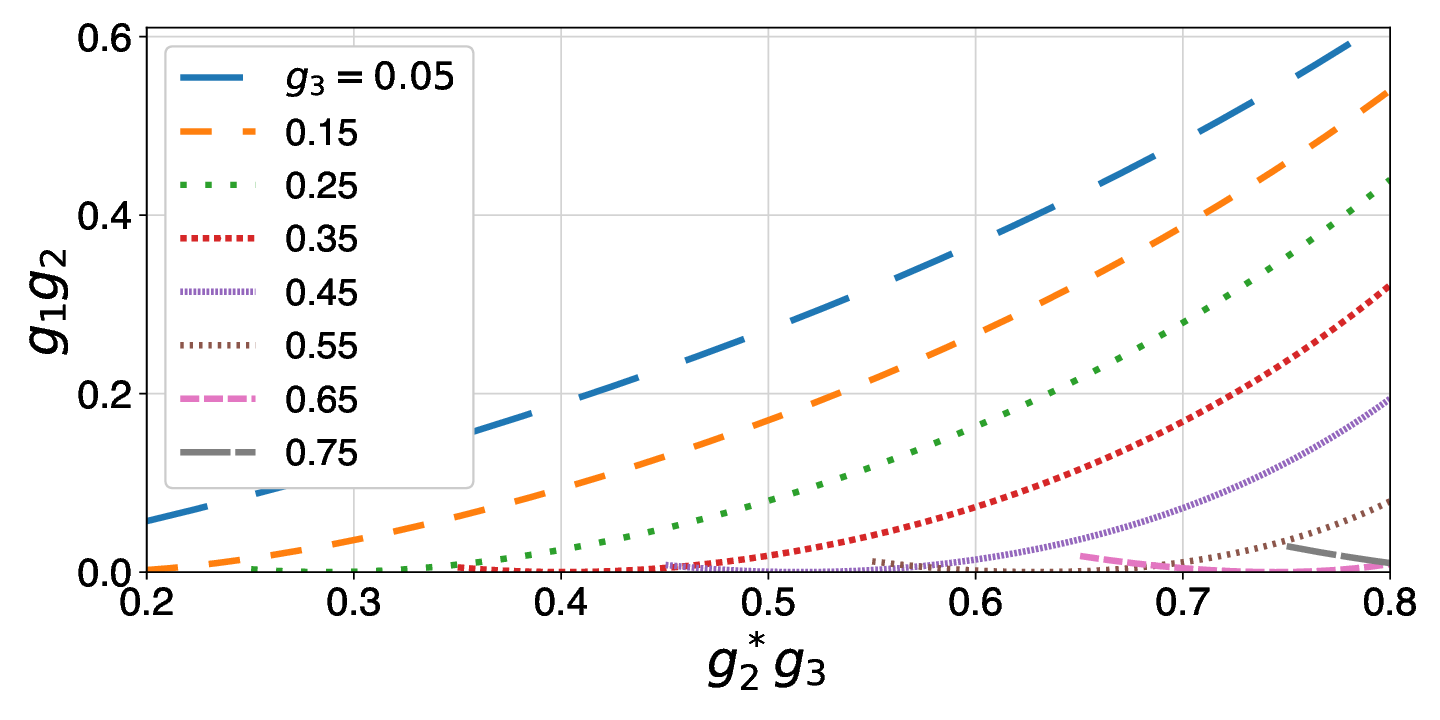}
        \caption{\centering Stability of the long biconcave cavity vs. stability of the short
        concave-convex cavity for different values of $g_{3}$}
        \label{fig:g2g3_vs_g1g2_short_convex}
    \end{subfigure}
    \hfill
    \centering
    \begin{subfigure}[h!]{.49\linewidth}
        \includegraphics[width=\textwidth]{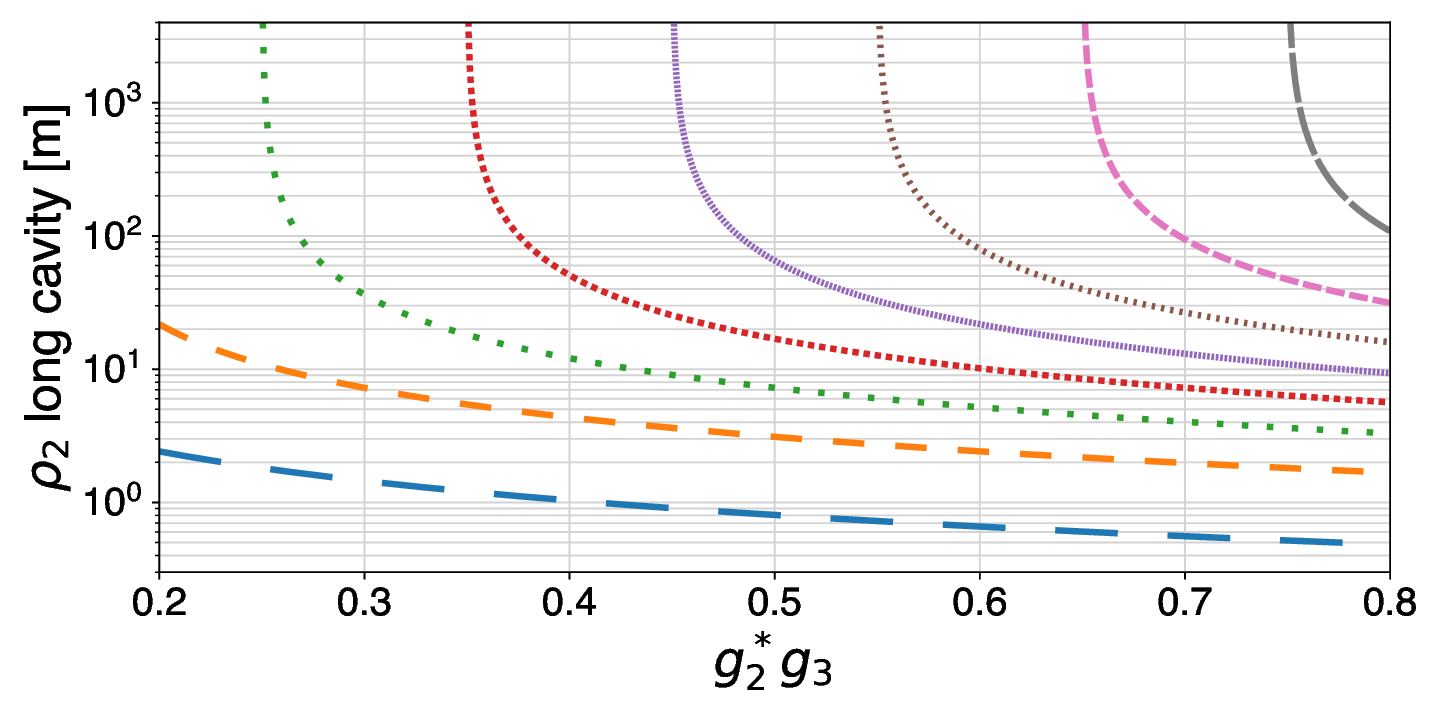}
        \caption{\centering Radius of curvature of the intermediate mirror seen from the long
        cavity in metres as a function of short cavity stability for different values
        of $g_{3}$.}
        \label{fig:R2_vs_g1g2}
    \end{subfigure}
    \caption{Geometrical properties of the long cavity as a function of the stability of the short cavity.}
    \label{fig:R2_vs_g1g2_bof}
\end{figure}

Let's just take as an example a long cavity with $L_1=L_\text{long}=\SI{50}{m}$ and a short cavity with $L_2=L_\text{short}=\SI{5}{m}$ and compute the stability factor $g_{1}g_{2}$ of the long cavity as a function of $g_{2}^{*}g_{3}$ for different values of $g_{3}$. The result is presented in Fig.~\ref{fig:g2g3_vs_g1g2_short_convex} and shows that the long cavity can be made stable if $g_3 \ll 1$, but both cavities cannot be made simultaneously very stable.

Fig.~\ref{fig:R2_vs_g1g2} shows the evolution of the intermediate mirror radius of curvature $\rho_2$ as a function of the stability of the short cavity. One sees that this radius of curvature increases rapidly with decreasing $g_{2}^{*}g_{3}$ whatever the value of $g_{3}$, or equivalently $\rho_{3}$. This means that the intermediate mirror tends to be flat, under most circumstances, a configuration which has already been described as not ideal.

\subsubsection{Short biconcave cavity}

In the case where a stable short biconcave symmetrical cavity is designed first, the Rayleigh range is of the order of its length $L_\text{short}$ according to Eq. (\ref{eq:rho2}). The beam propagating inside the long cavity can then accumulate a significant Gouy phase and this long cavity can thus be made stable. 

The same conclusion can be obtained by inspecting the $g_{2}^{*}g_{3}$ value of the long cavity.  As the long cavity is much longer than the Rayleigh range, then the radius of curvature of the end mirror, $\rho_{3}$, is only a little longer than the long cavity's length, which results in $g_{3}\gtrsim 0$. The reason for this large stability is mainly due to the fact that the intermediate radius of curvature $\rho_{2}^{*}$ is negative and shorter than the long cavity length, this results in a large $g_{2}^{*}$ value and as $g_{3}$ is positive their product
has a non negligible value. 

In practice, we start by building a short symmetrical biconcave cavity with a stability $g_{1}g_{2}=0.5$ using Eq. (\ref{eq:rho2}). We then build a long convex-concave cavity adapted to the beam geometry following using Eq. (\ref{eq:rho3}). In that case the only limitation of the long cavity length is the available mirror size for the end cavity mirror.

\subsection{Intermediate mirror designed as a lens-mirror component}
\label{sec:lens-mirror}

In the three-mirror cavity design we studied up to this point, the
anti-reflective backside of the intermediate mirror was a flat surface,
but this assumption is not mandatory, it is simply the common approach
to usual mirror design where a flat back surface lies outside
the area of interest, which is in front of the mirror. If we relax
this constraint and allow the anti-reflective surface to be curved,
we can build an intermediate mirror that acts simultaneously as a
curved mirror and as a lens. This extra degree of freedom may allow
to design stable three-mirror cavities, even with very large length
ratios. 

\begin{figure}[h!]
\centering
\includegraphics[width=0.795\textwidth]{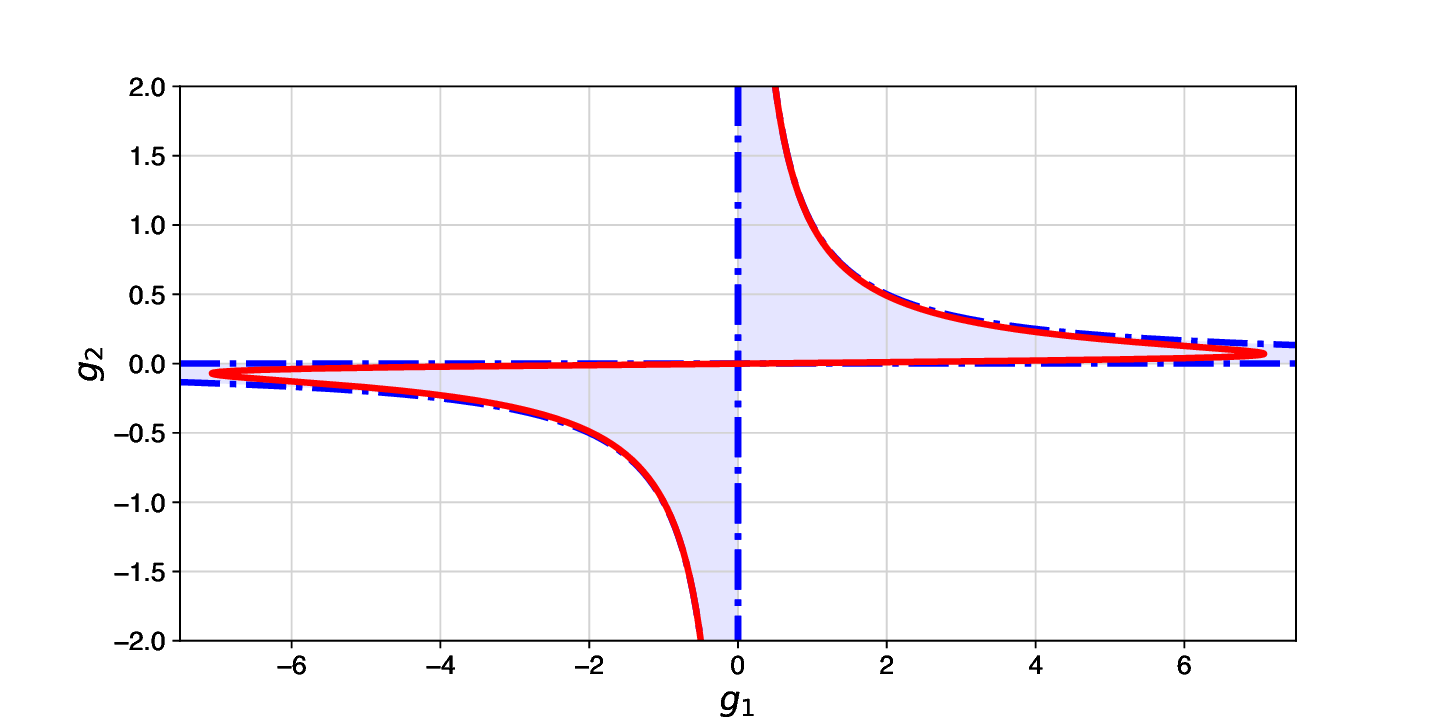}
\caption{Location of the $g_{1},g_{2}$ values, in red, for a continuous scan
of the radius of curvature of the anti-reflective side. The stability region is represented by filed area framed by blue dashed curves.\label{fig:Location-of-the_g1g2_path}}

\end{figure}

The possibility to add curvature to the anti-reflective side of the
intermediate mirror allows for changing the radius of curvature of the
beam at the intermediate mirror location as seen through the anti-reflection side
that lies, in the following example, in the first cavity. This
modifies the value of $g_{2}$ and for a given value of $L_{1}$ dictates
a value for $g_{1}$ and thus the stability factor $g_{1}g_{2}$ of
the first cavity.
As an example let us study the case of a main stable symmetric biconcave fifty-meter-long cavity with $g_{2}^{*}=g_{3}=1/\sqrt{2}$, coupled to a
shorter, five or three-meter cavity. The anti-reflective curved surface
lies in the short cavity and the long cavity, designed first, imposes
the beam geometry. For each value of the radius of curvature of the anti-reflective surface $\rho_{AR}$ we compute the value of $\rho_1$ which maximizes $g_1 g_2$.
\begin{figure}[h!]
\centering
\includegraphics[width=0.795\textwidth]{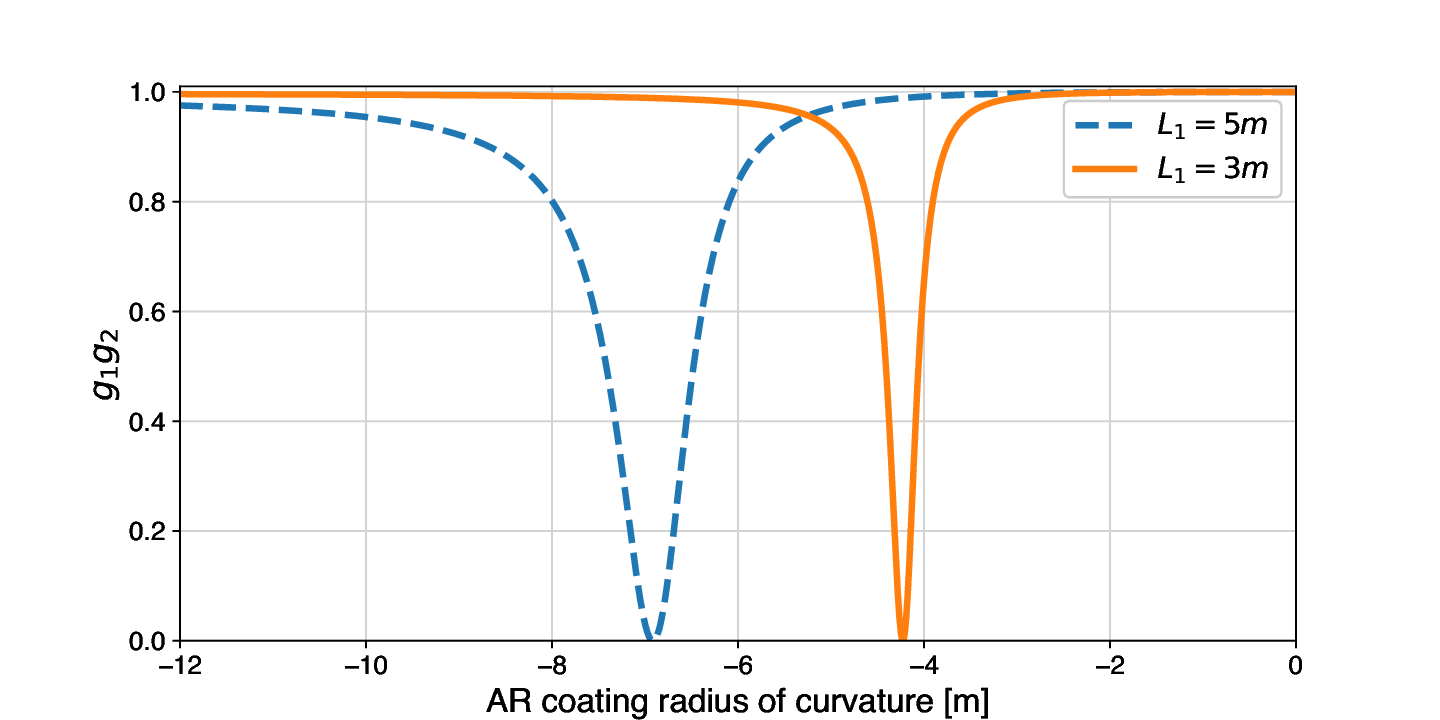}
\caption{Stability factor $g_{1}g_{2}$ of the short cavity of length $L_{1}$
as a function of the radius of curvature of the anti-reflective coating.
The long cavity length $L_{2}=50\:\mathrm{m}$. \label{fig:Stability-factor-vs-Rar}}
\end{figure}
The result, shown in Fig.~(\ref{fig:Stability-factor-vs-Rar}), demonstrates
that there exist some ranges of radii of curvature of the anti-reflective
surface that make the short cavity very stable. The extension of the
region of stability strongly depends on the ratio of the cavity lengths.
It is interesting to remark that the point where $g_{1}g_{2}=0$ corresponds
to a confocal cavity geometry. This can be understood as follows: by design we constrain the short cavity to be stable, even marginally.
This implies that the $g_{1}$, $g_{2}$ pair always lies within one
of the two stability regions of the cavity. In Fig.~\ref{fig:Location-of-the_g1g2_path} the two regions lie between the $g_{1}=0$ and $g_{2}=0$ lines and the two arches of the $g_{1}g_{2}=1$ hyperbola. As the radius of 
curvature of the anti-reflective coating is continuously changed, the
$g_{1},g_{2}$ parameter couple follows a path within these regions.
However for some value of the anti-reflective coating radius of curvature,
the sign of $g_{2}$ changes. This forces the sign of $g_{1}$ to
flip as well. The only common point of the two stability half regions
that allows for switching between two positive $g$ values and two negative
ones is precisely the point where $g_{1}=g_{2}=0$, which represents
a confocal cavity.

This result should not hide conceivable technical difficulties to
use such kind of optical components: aberrations or mode coupling
are much more enhanced by tilt and misalignment of curved surfaces
than flat ones, and the technical ability to realize optical surfaces
with radii of curvature precisely matched to requirements and sharing
the same optical axis is far from immediate. Another point that may
be of importance is that by changing the beam geometry in the first
cavity one modifies its Gouy phase through propagation, which may
lead to some issues in particular optical designs.

\section{Conclusion}

We have presented a few, non-exhaustive, original properties of three-mirror linear optical resonators, pointing their uniqueness and their limits in practical use. The coupling of two Fabry-Perot cavities sharing a common mirror cannot simply be reduced to the succession of two distinct resonators but that it creates a coupling of optical properties just like coupled mechanical or electrical resonators do. It results from this coupling a splitting of the resonance into a double-peak pattern. We have demonstrated that the reflectivities of mirrors and their relative spacing play different roles on the relative height between double-peak maxima, their position and their intrinsic width. Although it is possible to adjust the double peak shape in very different ways, special attention must be paid to asymmetrical configurations as the stability go inversely with the asymmetry for cavity with a flat central mirror. To some extent, these instabilities can be compensated by the use of curved anti-reflective surfaces but short biconcave cavity and long convex-concave one offers more stable solutions.

The study of these original properties may be valuable to fields of laser optics where shaping the behavior of optical resonators is an issue. In particular, the double-peak pattern could be controlled in real time by tuning the spacing between mirrors. This increase of controllability could be of interest for the optimization of future gravitational-wave detectors using squeezed states of light which has initially motivated our analysis.

\section{Acknowledgement}

The authors acknowledge the support of the French Agence Nationale de la Recherche (ANR), under Grant Nos. ANR-18-JSTQ-0002 (project QFilter). The authors thank Eleonora Capocasa for her helpful comments on this publication.

\printbibliography

\end{document}